\documentclass[aps,prd,nofootinbib,reprint,superscriptaddress]{revtex4-1}

\usepackage{blindtext}
\usepackage{amsfonts}
\usepackage{tensor}
\usepackage{graphicx}
\usepackage{pict2e}

\usepackage{balance}

\usepackage{amssymb, amsmath,environ}

\usepackage{color}

\newcommand{\vb}{\bar{v}}


\newcommand{\bse}{\begin{subequations}}
	\newcommand{\ese}{\end{subequations}}
\newcommand{\be}{\begin{equation}}
\newcommand{\ee}{\end{equation}}

\NewEnviron{eqsplit}{%
	\begin{equation}\begin{split}
	\BODY
	\end{split}\end{equation}
}

\makeatletter
\newcommand*\bigcdot{\mathpalette\bigcdot@{.5}}
\newcommand*\bigcdot@[2]{\mathbin{\vcenter{\hbox{\scalebox{#2}{$\m@th#1\bullet$}}}}}
\makeatother

\newcommand{\bea}{\begin{eqnarray}}
\newcommand{\eea}{\end{eqnarray}}
\newcommand{\ba}{\begin{array}}
	\newcommand{\ea}{\end{array}}

\newcommand{\nn}{{\nonumber}}

\newcommand{\Q}{\mathcal{Q}}
\newcommand{\A}{\mathcal{A}}
\newcommand{\PP}{\mathcal{P}}

\newcommand{\N}{\mathcal{N}}

\newcommand{\br}{\mathbf{r}}
\newcommand{\bk}{\mathbf{k}}
\newcommand{\C}{\mathcal{C}}
\newcommand{\Y}{\mathcal{Y}}

\newcommand{\OO}{\mathcal{O}}

\newcommand{\la}{\langle}
\newcommand{\ra}{\rangle}
\newcommand{\ve}{\varepsilon}

\begin{document}

\title{Non-Bessel-Gaussianity and Flow Harmonic Fine-Splitting}

%
%
%
%
%
%
%
%


\author{Hadi Mehrabpour}
\email[]{hadi.mehrabpour@ipm.ir}

\affiliation{Department  of  Physics,  Sharif  University  of  Technology,  P.O.  Box  11155-9161,  Tehran,  Iran}

\affiliation{School of Particles and Accelerators, Institute for Research in Fundamental Sciences (IPM), P.O. Box 19395-5531, Tehran, Iran}

\author{Seyed Farid Taghavi}
\email[]{s.f.taghavi@ipm.ir}

\affiliation{School of Particles and Accelerators, Institute for Research in Fundamental Sciences (IPM), P.O. Box 19395-5531, Tehran, Iran}

\affiliation{Physik Department, Technische Universit\"{a}t M\"{u}nchen, Munich, Germany}

\begin{abstract}

Both collision geometry and event-by-event fluctuations are encoded in the experimentally observed flow harmonic distribution $p(v_n)$ and $2k$-particle cumulants $c_n\{2k\}$. In the present study, we systematically connect these observables to each other by employing Gram-Charlier A series. We quantify the deviation of $p(v_n)$ from Bessel-Gaussianity in terms of flow harmonic fine-splitting. Subsequently, we show that the corrected Bessel-Gaussian distribution can fit the simulated data better than the Bessel-Gaussian distribution in the more peripheral collisions. Inspired by Gram-Charlier A series, we introduce a new set of cumulants $q_n\{2k\}$ that are more natural to study distributions near Bessel-Gaussian. These new cumulants are obtained from $c_n\{2k\}$ where the collision geometry effect is extracted from it. By exploiting $q_2\{2k\}$, we introduce a new set of estimators for averaged ellipticity $\vb_2$ which are more accurate compared to $v_2\{2k\}$ for $k>1$.  As another application of $q_2\{2k\}$, we show we are able to restrict the phase space of $v_2\{4\}$, $v_2\{6\}$ and $v_2\{8\}$ by demanding the consistency of $\vb_2$ and $v_2\{2k\}$ with $q_2\{2k\}$ equation.  The allowed phase space is a region such that $v_2\{4\}-v_2\{6\}\gtrsim 0$ and $12 v_2\{6\}-11v_2\{8\}-v_2\{4\}\gtrsim 0$, which is compatible with the experimental observations.

\end{abstract}

\maketitle

\section{Introduction}

It is a well-established picture that the produced matter in the heavy ion experiment has collective behavior. Based on this picture, the initial energy density manifests itself in the final particle momentum distribution. Accordingly, as the main collectivity consequence, the final particle momentum distribution is extensively studied by different experimental groups in the past years. As a matter of fact, the experimental groups at Relativistic Heavy Ion Collider (RHIC) and Large Hadron Collider (LHC) measure the flow harmonics $v_n$ \cite{Ackermann:2000tr,Lacey:2001va,Park:2001gm,Aamodt:2010pa,ALICE:2011ab,Chatrchyan:2012ta,ATLAS:2011ah,Aad:2014vba},
the coefficients of the momentum distribution Fourier expansion in the azimuthal direction \cite{Ollitrault:1992bk,Barrette:1994xr,Barrette:1996rs}.  All these observations can be explained by several models based on the collective picture.


Finding the flow harmonics is not straightforward,  because the reaction plane angle (the angle between the orientation of the impact parameter and a reference frame) is not under control experimentally.  Additionally, the Fourier coefficients cannot be found reliably due to the low statistic in a single event. These enforce us to use an analysis more sophisticated than a Fourier analysis. There are several methods to find the flow harmonics experimentally, namely Event-Plane method \cite{Poskanzer:1998yz}, multiparticle correlation functions \cite{Borghini:2001vi,Borghini:2000sa} and Lee-Yang zeros \cite{Bhalerao:2003yq,Bhalerao:2003xf}. The most recent technique to find the flow harmonics is using the distribution of flow harmonic $p(v_n)$. This distribution  has been obtained experimentally by employing the unfolding technique \cite{Jia:2013tja,Aad:2013xma}.  

It is known that the initial shape of the energy density depends on the geometry of the collision and the quantum fluctuations at the initial state. As a result, the observed flow harmonics fluctuate event by event even if we fix the initial geometry of the collision. In fact, the event-by-event fluctuations are encoded in  $p(v_n)$ and experimentally observed flow harmonics as well. 
It is worthwhile to mention that the observed event-by-event fluctuations are a reflection of the initial state fluctuations entangled with the modifications during different stages of the matter evolution, namely  collective expansion and the hadronization. For that reason, exploring the exact interpretation of the flow harmonics is crucial to understand the contribution of each stage of the evolution on the fluctuations. Moreover, there is no well-established picture for the initial state of the heavy ion collision so far. The interpretations of the observed quantities contain information about the initial state. This information can shed light upon the heavy ion initial state models too.

According to the theoretical studies, flow harmonic $v_n\{2k\}$ obtained from  $2k$-particle correlation functions are different, and their difference is sourced by non-flow effects \cite{Borghini:2000sa,Borghini:2001vi} and event-by-event fluctuations as well \cite{Miller:2003kd}. We should say that experimental observations show that $v_2\{2\}$ is considerably larger than $v_2\{4\}$, $v_2\{6\}$ and $v_2\{8\}$. Additionally, all the ratios $v_2\{6\}/v_2\{4\}$, $v_2\{8\}/v_2\{4\}$ and $v_2\{8\}/v_2\{6\}$ are different from unity \cite{Sirunyan:2017fts,Acharya:2018lmh}. Alternatively, the distribution $p(v_n)$ is approximated by Bessel-Gaussian distribution (corresponds to a Gaussian distribution for $v_n$ fluctuations) as a simple model \cite{Barrette:1994xr,Voloshin:2007pc}. Based on this model, the difference between $v_2\{2\}$ and $v_2\{4\}$ is related to the width of the $v_2$ fluctuations. However, this model cannot explain the difference between other $v_2\{2k\}$.

In the past years, several interesting studies about the non-Gaussian $v_n$ fluctuations have been done \cite{Bhalerao:2011bp,Yan:2013laa,Yan:2014afa,Gronqvist:2016hym,Giacalone:2016eyu,Abbasi:2017ajp}. Specifically, it has been shown in Ref.~\cite{Giacalone:2016eyu} that the skewness of $v_2$ fluctuations is related to the difference $v_2\{4\}-v_2\{6\}$. Also, the connection between kurtosis of $v_3$ fluctuations and the ratio $v_3\{4\}/v_3\{2\}$ has been studied in Ref.~\cite{Abbasi:2017ajp}. It is  worth noting that the deviation of $v_n$ fluctuations from Gaussian distribution immediately leads to the deviation of $p(v_n)$ from Bessel-Gaussian distribution. In \cite{Jia:2014pza}, the quantities $v_n\{4\}-v_n\{6\}$ and $v_n\{6\}-v_n\{8\}$ for a generic narrow distribution is computed.

In the present work, we will introduce a systematic method to connect $v_n\{2k\}$ to distribution $p(v_n)$. In Sec.~\ref{secII}, we have an overview of the known concepts of cumulants, flow harmonic distributions and their relation with the averaged flow harmonics $\vb_n$. Sec.~\ref{seIII} is dedicated to the Gram-Charlier A series in which we find an approximated flow harmonic distribution in terms of $c_n\{2k\}$. Specifically for the second harmonics, we show that the deviation of $p(v_2)$ from Bessel-Gaussianity is quantified by the fine-splitting $v_2\{2k\}-v_2\{2\ell\}$ where $k,\ell \geq2$ and $k\neq \ell$. These studies guide us to define a new set of cumulants $q_n\{2k\}$ where they depend on the event-by-event fluctuations only. In Sec.~\ref{findAverage}, we use the  new cumulants to introduce more accurate estimations for the average ellipticity. As another application of new cumulants, we use $q_2\{2k\}$ to constrain the $v_2\{4\}$, $v_2\{6\}$ and $v_2\{8\}$ phase space in Sec.~\ref{secV}. We show that the phase space is restricted to a domain that  $v_2\{4\}-v_2\{6\}\gtrsim 0$\footnote{In Ref.~\cite{Giacalone:2016eyu}, the constraint $v_2\{4\}>v_2\{6\}$ is deduced from the initial eccentricity, and the fact that the initial eccentricity is bounded to a unit circle.  } and $12 v_2\{6\}-11v_2\{8\}-v_2\{4\}\gtrsim 0$. We present the conclusion in Sec.~\ref{secVI}. The supplementary materials can be found in the appendices. We would like to emphasize that in the Appendix~\ref{AppC}, we found a one-dimensional distribution for $p(v_n)$ which is different from that mentioned in Sec.~\ref{seIII}. Additionally, an interesting connection between $p(v_n)$ expansion in terms of cumulants $c_n\{2k\}$ and a relatively new concept of the multiple orthogonal polynomials in mathematics is presented in the Appendix~\ref{appD}.

\section{Flow Harmonic Distributions and $2k$-Particle Cumulants}\label{secII}

This section is devoted to an overview of already known concepts of cumulant and its application to study the collectivity in the heavy ion physics. We present this overview to smoothly move forward to the flow harmonic distribution and its deviation from Bessel-Gaussianity.

\subsection{Correlation Functions vs. Distribution}\label{IIA}

According to the collective picture in the heavy ion experiments, the final particle momentum distribution is a consequence of the initial state after a collective evolution. In order to study this picture quantitatively, the initial anisotropies and flow harmonics  are used extensively to quantify the initial energy density and final momentum distribution.

The initial energy (or entropy) density of a single event can be written in terms of the initial anisotropies, namely the ellipticity and triangularity. Specifically, ellipticity and triangularity (shown by $\ve_2$ and $\ve_3$, respectively) are cumulants of the distribution indicating how much it is deviated from a two dimensional rotationally symmetric Gaussian distribution \cite{Teaney:2010vd}.

The final momentum distribution is studied by its Fourier expansion in the azimuthal direction,
\bea\label{flowFourier}
\frac{1}{N}\frac{dN}{d\phi}=\frac{1}{2\pi}\left[1+\sum_{n=1}^{\infty} 2 v_n \cos n(\phi-\psi_n)\right].
\eea
In the above, $v_n$ and $\psi_n$ are unknown parameters that can be found easily via $\hat{v}_n\equiv v_n e^{in\psi_n}=\la e^{i n \phi} \ra$. Here, the averaging is obtained by using the distribution $\frac{1}{N}\frac{dN}{d\phi}$ in a given event. The parameter $\hat{v}_n$ is called flow harmonic. Instead of complex form, we occasionally use the flow harmonics in Cartesian coordinates,
\bea
v_{n,x}=v_n\cos n\psi_n,\qquad v_{n,y}=v_n\sin n\psi_n.
\eea

Low multiplicity in a single event, randomness of the reaction plane angle and non-flow effects are the main challenges in extracting the experimental values of flow harmonics. By defeating these experimental challenges, one would be able to find the distribution $p(v_{n,x},v_{n,y})$ for each centrality class which is a manifestation of the event-by-event fluctuations.  In finding $p(v_{n,x},v_{n,y})$, we rotate the system at each single event such that the reaction plane angle $\Phi_{\text{RP}}$ is set to be zero. In this case, the \textit{averaged ellipticity} $\vb_2\equiv \la v_{2,x} \ra $ is a manifestation of the geometrical initial ellipticity for events in a given centrality class irrespective of the fluctuations. In general, we are able to define \textit{averaged flow harmonic} $\vb_n\equiv \la v_{n,x} \ra $ too. In odd harmonics, however, this average would be zero for spherical ion collisions with same sizes.

A practical way to study a distribution function $p(\mathbf{r})$ is using two dimensional cumulants, where $\mathbf{r}$ stands for a generic two dimensional random variable. Consider $G(\mathbf{k})$ as the characteristic function of the probability distribution $p(\mathbf{r})$. The characteristic function is, in fact, the Fourier transformed $p(\mathbf{r})$,
\bea\label{generatingFu}
G(\mathbf{k})\equiv \la e^{i \mathbf{r}\cdot \mathbf{k}} \ra = \int dxdy\,e^{i \mathbf{r}\cdot \mathbf{k}} p(\mathbf{r}).
\eea
Consequently, we can define the cumulative function as $\mathcal{C}(\mathbf{k})=\log G(\mathbf{k})$. Two dimensional cumulants are obtained from
\begin{equation}\label{2DCumuls}
\begin{aligned}
\log\la e^{i \mathbf{r}\cdot \mathbf{k}} \ra&=\sum_{m,n}\frac{i^{m+n}k_x^m k_y^n \A_{mn}}{m! n!}\\
&=\sum_{m=0}^{\infty}\sum_{n=-\infty}^{\infty} \frac{i^m k^m e^{in\varphi_k} \;\C_{m,n}}{m!},
\end{aligned}
\end{equation}
where $\A_{mn}$ and $\C_{m,n}$ are the 2D cumulants in Cartesian and polar coordinates, respectively. The two dimensional cumulants have  already been used to quantify the initial energy density shape by Teaney and Yan in Ref.~\cite{Teaney:2010vd}. However, we use the two dimensional cumulants to study flow harmonic distributions in the present study\footnote{In Ref.~\cite{Teaney:2010vd}, $\A_{mn}$ has been shown by $W_{n,ab}$ ($n=1,2,\ldots$ and $a,b\in\{x,y\}$). Also $\C_{mn}$ has been shown by $W_{0,n}$, $W^s_{0,n}$ and $W^c_{0,n}$. }. In the second line in the above, we used $k=\sqrt{k_x^2+k_y^2},\,\varphi_k=\text{atan2}(k_x,k_y)$. In Appendix \ref{AppA}, we study 2D cumulants in these two coordinates  and their relations in more details.

The random $\Phi_{\text{RP}}$ rotates the point $(v_{n,x},v_{n,y})$ with a random phase in the range $[0,2\pi)$ at each event. As a result, the distribution $p(v_{n,x},v_{n,y})$ is replaced by a $\tilde{p}(v_{n,x},v_{n,y})$ which is a rotationally symmetric distribution\footnote{In general $p(v_{n,x},v_{n,y})$ is not rotationally symmetric. For instance $p(v_{2,x},v_{2,y})$ has not this  symmetry in non-central collision of spherically symmetric ions while $p(v_{3,x},v_{3,y})$ has it for the same collisions. }. 
We should say that the distribution $\tilde{p}(v_{n,x},v_{n,y})$  is experimentally accessible. In Ref.~\cite{Aad:2013xma}, the unfolding technique has been used by ATLAS collaboration 
to remove the statistical uncertainty (sourced by low statistics at each event) and non-flow effects from a distribution of  ``observed'' flow harmonics $(v_{n,x}^{\text{obs}},v_{n,y}^{\text{obs}})$.
In this case, the only unknown parameter to find an accurate $p(v_{n,x},v_{n,y})$ is the reaction plane angle.
Since there is no information in the azimuthal direction of $\tilde{p}(v_{n,x},v_{n,y})$, we can simply average out this direction to find a one dimensional distribution\footnote{We simply use the notation $\varphi\equiv n\psi_n.$},
\bea\label{1DDist}
p(v_n)=v_n \int_{0}^{2\pi}d\varphi\, p(v_n \cos\varphi,v_n \sin\varphi).
\eea
Note that we can interchangeably use  $\tilde{p}(v_{n,x},v_{n,y})$ or $p(v_{n,x}$ $,v_{n,y})$ in the above because obviously the effect of the random reaction plane angle and the azimuthal averaging are the same.

In polar coordinates, we have $\tilde{p}(v_{n,x},v_{n,y})\equiv \tilde{p}(v_n)$. As a result, the characteristic function of the distribution $\tilde{p}(v_{n,x},v_{n,y})$ in polar coordinates is given by
\begin{equation}\label{2DCharto1DChar}
\begin{aligned}
\la & e^{i v_n k\cos(\varphi-\varphi_k)} \ra_{\text{2D}}=\\
&\int_0^{\infty}\int_0^{2\pi} v_n d v_nd\varphi\; \tilde{p}(v_n)\;e^{ i v_n k\cos(\varphi-\varphi_k)}\\
&= \int_0^{\infty} dv_n\,p(v_n) J_0(k v_n)=\la J_0(k v_n) \ra_{\text{1D}},
\end{aligned}
\end{equation}
where we used Eq.~\eqref{1DDist} in the above. Here, $J_0(x)$ is the Bessel function of the first kind. Also, $\la \cdots\ra_{\text{2D}}$ means averaging with respect to $\tilde{p}(v_{n,x},v_{n,y})$ while $\la \cdots\ra_{\text{1D}}$ specifies the averaging with respect to $p(v_n)$. The Eq.~\eqref{2DCharto1DChar} indicates that we can study the \textit{radial} distribution $p(v_n)$ instead of $\tilde{p}(v_{n,x},v_{n,y})$ if we use $G(k)=\la J_0(k v_n) \ra$  as the characteristic function of $p(v_n)$\footnote{We ignore the subscript 1D or 2D when it is not ambiguous.}.

The cumulants of $p(v_n)$ can be found by expanding the  cumulative function $\log \la J_0 (k v_n)\ra $ in terms of $ik$. The coefficients of $ik$ in this expansion (up to some convenient constants) are the desired cumulants,
\bea\label{cnCumul}
\log \la J_0 (k v_n)\ra = \sum_{m=1}\frac{(ik)^{2m} c_n\{2m\}}{4^m(m!)^2}.
\eea
where $c_n\{2m\}$ are those obtained from  $2k$-particle correlation functions \cite{Borghini:2000sa,Borghini:2001vi},
\begin{subequations}\label{1DnPartCumul}
	\begin{eqnarray}
	c_n\{2\}&=&\la v_n^2 \ra, \label{1DnPartCumulA}\\
	c_n\{4\}&=&\la v_n^4 \ra-2\la v_n^2 \ra^2, \label{1DnPartCumulB}\\
	c_n\{6\}&=&\la v_n^6 \ra-9\la v_n^4 \ra\la v_n^2 \ra+12\la v_n^2 \ra^3, \label{1DnPartCumulC}\\
	c_n\{8\}&=&\la v_n^8 \ra-16\la v_n^6 \ra\la v_n^2 \ra-18\la v_n^4 \ra^2, \label{1DnPartCumulD}\\
	&&+144\la v_n^4 \ra\la v_n^2\ra^2-144\la v_n^2 \ra^4.\nonumber
	\end{eqnarray}
\end{subequations}

The cumulants $c_n\{2k\}\propto v_n^{2k}\{2k\}$ (see Eq.~\eqref{2kFlowHArmonic}) are indicating the characteristics of the distribution $p(v_n)$ while the cumulants $\A_{mn}$ (or equivalently $\C_{m,n}$) in Eq.~\eqref{2DCumuls} are shown the characteristics of $p(v_{n,x},v_{n,y})$. The interconnection between $v_n\{2k\}$ and $\A_{mn}$ have been studied previously in the literature. Point out that in order to define $p(v_{n,x},v_{n,y})$, we considered $\Phi_{\text{RP}}=0$ for all events. In this case, the cumulant $\A_{30}=\la(v_{n,x}$  $-\la v_{n,x}\ra)^3 \ra$
is related to the skewness of $p(v_{n,x},v_{n,y})$. For the case $n=2$, this quantity is found for the first time in Ref.~\cite{Giacalone:2016eyu}, and it is argued that $\A_{30}\propto v_2\{4\}-v_2\{6\}$. In other words, $\A_{30}$ is related to the \textit{fine-splitting} between $v_2\{4\}$ and $v_2\{6\}$. In Ref.~\cite{Abbasi:2017ajp}, the kurtosis of $p(v_{3,x},v_{3,y})$ in the radial direction has also been calculated, and it is shown it is proportional to $v_3^4\{4\}$\footnote{The coefficients of the proportionality in both skewness and radial kurtosis are also functions of $v_n\{2k\}$. We should note that the skewness (radial kurtosis) vanishs if $v_2\{4\}-v_2\{6\}$ ($v_3\{4\}$) is equal to zero (see Refs.~\cite{Giacalone:2016eyu,Abbasi:2017ajp} for the details).  }.

One may wonder whether the distribution $p(v_n)$ contains more information than $c_n\{2k\}$ because the odd moments of $p(v_n)$ are absent in the definitions of $c_n\{2k\}$ \cite{Jia:2014pza}. However, it is worth mentioning that the moments can be found by expanding the characteristic function $G(k)$ in terms of $ik$,
\bea\label{charaSeris}
G(k)=\la J_0 (k v_n)\ra=1+\frac{\la v_n^2 \ra }{4} k^2+\frac{\la v_n^4 \ra}{64} k^4+\cdots,
\eea 
for the radial distributions like $p(v_n)$. Since the above series is convergent\footnote{It is an important question whether is it possible to determine $p(v_n)$ uniquely from its moments \cite{Stoyanov} (see also Ref.~\cite{Bilandzic:2013kga})? Answering to this question is beyond the scope of the present work. Here we assume that $p(v_n)$ is \textit{M-determinate} which means we can find it from its moments $\la v_n^{2q}\ra $ in principle. }, we can find the characteristic function $G(k)$ by knowing the moments $\la v_n^{2k} \ra$.
Having characteristic function in hand, we immediately find $p(v_n)$ by inversing the last line in the Eq.~\eqref{2DCharto1DChar}\footnote{We use the orthogonality relation $\int_0^{\infty}k\,J_{\alpha}( k r)J_{\alpha}( k r') \,dk=\delta(r-r')/r$.},
\bea
p(v_n)=v_n \int_0^{\infty}k\, dk \,J_o(k v_n) G(k).
\eea
It means that by assuming the convergence of the series in Eq.~\eqref{charaSeris} the distribution $p(v_n)$ can be found completely by using only even moments.

Equivalently, we can use the following argument: for the distribution  $\tilde{p}(v_{n,x},v_{n,y})$, one simply finds the only non-vanishing moments are $\la v_{n_x}^{2k} v_{n_y}^{2\ell} \ra$. It means that in the polar coordinates only $\la v_n^{2(k+\ell)}\ra$ are present. Additionally, by finding the two dimensional cumulants of $\tilde{p}(v_{n,x},v_{n,y})$ in polar coordinates $\C_{m,n}$ (Eq.~\eqref{2DCumuls}),  we obtain that the only non-zero cumulants are $\C_{2k,0}\propto c_n\{2k\}$ (see Appendix~\ref{AppA}).

As a result, in the presence of random reaction plane angle, $c_n\{2k\}$'s are all we can learn from the original $p(v_{n,x},v_{n,y})$, whether we use $2k$-particle correlation functions or obtain it from the unfolded distribution $p(v_n)$ in principle. However, we should say that the efficiency of the two methods in removing single event statistical uncertainty and non-flow effects could be different which leads to different results in practice.

Furthermore, the whole information about the fluctuations are not encoded in $p(v_{n,x},v_{n,y})$. In fact, the most general form of the fluctuations are encoded in a distribution as  $p(v_{1,x},v_{1,y},v_{2,x},v_{2,y},\ldots)$. It is worth mentioning that the symmetric cumulants, which have been introduced in Ref.~\cite{Bilandzic:2013kga} and have been measured by ALICE collaboration \cite{ALICE:2016kpq}, are non-vanishing. Additionally, the event-plane correlations (which are related to the moments $\la \hat{v}_m^q (\hat{v}^*_{n})^{q\,m/n} \ra $) have been obtained by the ATLAS collaboration \cite{Jia:2012sa,Aad:2014fla}. They are non-zero too.
These measurements
indicate that $p(v_{1,x},v_{1,y},v_{2,x},v_{2,y},\ldots)$ cannot be written as $\prod_n p(v_{n,x},v_{n,y})$. In the present work, however, we do not focus on the joint distribution and leave this topic for the studies in the future. Let us point out that moving forward to find a generic form for the moments of the flow harmonic distribution was already done in Ref.~\cite{Bilandzic:2014qga}.

\subsection{Approximated Averaged Ellipticity}\label{firstApproxSubSec}

A question arises now: how much information is encoded in $p(v_n)$ from the original $p(v_{n,x},v_{n,y})$? In order to answer this question, we first focus on $n=3$. Unless there is no \textit{net} triangularity for spherical ion collisions, the non-zero triangularity at each event comes from the fluctuations. Hence, we have $\vb_3=0$ for such an experiment. In this case, the event-by-event randomness of $\Phi_{\text{RP}}$ is similar to the rotation of the triangular symmetry plane due to the event-by-event fluctuations.
It means that $p(v_{3,x},v_{3,y})$  itself is rotationally symmetric, and the main features of  $p(v_{3,x},v_{3,y})$ and $\tilde{p}(v_{3,x},v_{3,y})$ are same. As a consequence, $p(v_3)$ or equivalently $c_3\{2k\}$ can uniquely reproduce the main features of $p(v_{3,x},v_{3,y})$. 

However, it is not the case for $n=2$ due to the non-zero averaged ellipticity $\vb_2$. The distribution $p(v_{2,x},v_{2,y})$ is not rotationally symmetric  and reshuffling $(v_{2,x},v_{2,y})$ leads to information loss from the original $p(v_{2,x},v_{2,y})$.
Therefore, there is at least some information in $p(v_{2,x},v_{2,y})$ that we cannot obtain it from $p(v_2)$ or $c_2\{2k\}$. Nevertheless, it is still possible to find some features of $p(v_{2,x},v_{2,y})$ approximately. For instance, we mentioned earlier in this section that the skewness of this distribution in the $v_{2,x}$ direction  is proportional to $v_2\{4\}-v_2\{6\}$.

The other important feature of $p(v_{2,x},v_{2,y})$ is $\vb_2$ which is not obvious how we can find from $c_2\{2k\}$. In fact, we are able to approximately find $\vb_2$ in terms of $c_2\{2k\}$ by approximating $p(v_{2,x},v_{2,y})$. The most trivial approximation is a two dimensional Dirac delta function located at $(\vb_2,0)$,
$$p(v_{2,x},v_{2,y})\simeq \delta(v_{2,x}-\vb_2,v_{2,y}).$$
This corresponds to the case that there are no fluctuations, and the only source for ellipticity is coming from an ideally elliptic initial geometry. Considering Eq.~\eqref{1DDist}, the moments $\la v_2^{2q} \ra$ can be easily obtained as follows
\bea\label{deltaDist}
\la v_2^{2q} \ra&=&\int d v_{2,x} d v_{2,y} (v_{2,x}^2+v_{2,x}^2)^{q} \delta(v_{2,x}-\vb_2,v_{2,y}\nn\\
&=& \vb_2^{2q}.
\eea
Now by using Eq.~\eqref{1DnPartCumul}, we find $c_2\{2\}=\vb_2^2$, $c_2\{4\}=-\vb_2^4$, $c_2\{6\}=4\vb_2^6$, etc. It is common in the literature to show the averaged ellipticity $\vb_2$ which is approximated by $c_2\{2k\}$ as $v_2\{2k\}$. Note that the quantity $v_2\{2k\}$  is \textit{defined} for the case that the flow harmonic distribution is considered as a delta function.  Furthermore,  we can assume an ideal case that for any harmonics the distribution $p(v_{n,x},v_{n,x})$ has a sharp and clean peak around $\vb_n$. By the mentioned assumption, we have \cite{Borghini:2001vi},
\begin{equation}\label{2kFlowHArmonic}
\begin{aligned}
v_n^2\{2\}&=c_n\{2\},\\
v_n^4\{4\}&=-c_n\{4\},\\
v_n^6\{6\}&=c_n\{6\}/4,\\
v_n^8\{8\}&=-c_n\{8\}/33.
\end{aligned}
\end{equation}
Nevertheless, we know that $\vb_n$ can be non-zero for odd $n$ when the collided ions are not spherical or have different sizes\footnote{The assumptions which have been used in Ref.~\cite{Borghini:2001vi} to find Eq.~\eqref{2kFlowHArmonic} are exactly equivalent to considering $p(v_{n,x},v_{n,y})$ as a delta function.  }.

The delta function approximation for $p(v_{2,x},v_{2,y})$ is not compatible with the experimental observation. In this case, we have
\bea\label{deltaVn}
v_2\{2\}=v_2\{4\}=\cdots=\vb_2,
\eea
by definition, while different $v_2\{2k\}$ have different values based on the experimental observation. Specifically, the difference between $v_2\{2\}$ and other $v_2\{2k\}$ for $k>1$ is considerably large \cite{Sirunyan:2017fts,Acharya:2018lmh}.

We can improve the previous approximation by replacing the delta function with a Gaussian distribution. In this case, we model the fluctuations by the width of the Gaussian distribution. Let us assume that $p(v_{2,x},v_{2,y})\simeq\N(v_{2,x},v_{2,y})$ where $\N(v_{2,x},v_{2,y})$ is a two dimensional Gaussian distribution located at $(\vb_2,0)$\footnote{We consider a reasonable assumption that the widths of the Gaussian distribution in the $v_{2,x}$ and $v_{2,y}$ directions are same. },
\bea\label{2DGauss}
\N(v_{2,x},v_{2,y})=\frac{1}{2\pi \sigma^2} e^{-\frac{(v_{2,x}-\vb_2)^2+v_{2,y}^2}{2\sigma^2}}.
\eea
Using above and Eq.~\eqref{1DDist}, one can simply find $p(v_2)=\text{BG}(v_2;\vb_2)$  where $\text{BG}(v_2;\vb_2)$ is the well-known Bessel Gaussian distribution \cite{Barrette:1994xr,Voloshin:2007pc},
\bea\label{BG}
\text{BG}(v_2;\vb_2)=\left(\frac{v_2}{\sigma^2}\right) I_0\left(\frac{v_2 \vb_2}{\sigma^2}\right) e^{-\frac{v_2^2+\vb_2^2}{2\sigma^2}}.
\eea
Here, $I_0(x)$ is the modified Bessel function of the first kind. Now, we are able to find the moments $\la v_2^{2q} \ra$ by using this approximated $p(v_2)$. According to the relations in Eq.~\eqref{1DnPartCumul}, we find
\begin{equation}\label{BGCUMUls}
\begin{aligned}
v_2\{2\}&=\sqrt{\vb_2^2+2\sigma^2}\\
v_2\{4\}&=v_n\{6\}=v_n\{8\}=\cdots =\vb_2,
\end{aligned}
\end{equation}
where we used the notation $v_2\{2k\}$ introduced in Eq.~\eqref{2kFlowHArmonic}. This result explains the large difference between $v_2\{2\}$ and $v_2\{2k\}$ for $k>1$. In fact, the presence of fluctuations is responsible for it. This description for the difference between $v_2\{2\}$ and other flow harmonics was argued first in Ref.~\cite{Voloshin:2007pc}. It is found that the splitting between $v_2\{2\}$ and other flow harmonics contains information from the two dimensional distribution \cite{Voloshin:2007pc}. 

The above two examples bring us to the following remarks:
\begin{itemize}
	\item  	In order to relate $\vb_n$ to $c_n\{2k\}$,  one needs to  estimate the  shape of $p(v_n)$ where $\vb_n$ is implemented in this estimation explicitly. We show this estimated distribution by $p(v_n;\vb_n)$.
	\item One can check the accuracy of the estimated distribution by studying the fine-splitting $v_n\{2k\}-v_n\{2\ell\}$ and comparing it with the experimental data.
\end{itemize}
We should say that the first remark is very strong and we can estimate $\vb_n$ by a weaker condition. Obviously, if we estimate only one moment or cumulant of $p(v_n)$  as a function of $\vb_n$,  in principle, we can estimate $\vb_n$ by comparing the estimated moment or cumulant with the experimental data. But the question is how to introduce such a reasonable estimation practically.
In the following sections, we introduce a method to estimate $\vb_n$ from a minimum information of  $p(v_n)$.  
One notes that $\vb_2=v_2\{4\}=\cdots $ is true only if we approximate  $p(v_2)$ by Bessel-Gaussian distribution. In the next section, we find an approximated distribution around Bessel-Gaussianity.

\section{Radial-Gram-Charlier Distribution and New Cumulants}\label{seIII}

In Sec.~\ref{firstApproxSubSec}, we argued that 
the quantity $\vb_n$, which is truly related to the geometric features of the collision,  can be obtained by estimating a function for $p(v_{n,x},v_{n,y})$. We observed that the Dirac delta function choice for $p(v_{n,x},v_{n,y})$ leads to $\vb_n=v_n\{2k\}$ for $k>0$, while by assuming the distribution as a 2D Gaussian located at $(\vb_n,0)$ (the Bessel-Gaussian in one dimension) we find $\vb_n=v_n\{2k\}$ for $k>1$. One notes that in modeling the flow harmonic distribution by delta function or Gaussian distribution, the parameter $\vb_n$ is an unfixed parameter which is eventually related to the $v_n\{2k\}$. In any case, the experimental observation indicates that $v_n\{2k\}$ are split, therefore, the above two models are not accurate enough.

Instead of modeling $p(v_{n,x},v_{n,y})$, we will try to model $p(v_n)$ with an unfixed parameter $\vb_n$, namely $p(v_n;\vb_n)$.  In this section, we introduce a series for this distribution such that the leading term in this expansion is the Bessel-Gaussian distribution. The expansion coefficients would be a new set of cumulants that specifies the deviation of the distribution from the Bessel-Gaussianity. In fact, by using these cumulants, we would be able to model $p(v_n;\vb_n)$ more systematically.

It is well-known that a given distribution can be approximated by Gram-Charlier A series which approximate the distribution in terms of its cumulants (see Appendix~\ref{AppB}). Here we use this concept to find an approximation for $p(v_n;\vb_n)$ in term of cumulants $c_n\{2k\}$. One of the formal methods of finding the Gram-Charlier A series is using orthogonal polynomials. In addition to this well-known method, we will introduce an alternative method which is more practical for finding the series of a one dimensional $p(v_n;\vb_n)$  around the Bessel-Gaussian distribution. 

\subsection{Gram-Charlier A series: 1D Distribution with Support  $\mathbb{R}$}\label{IIIA}

Before finding the approximated distribution around Bessel-Gaussian, let us practice the alternative method of finding Gram-Charlier A series by applying it to a one-dimensional distribution $p(x)$ with support $(-\infty,\infty)$\footnote{An standard method for finding the Gram-Charlier A series of $p(x)$ is reviewed in Appendix~\ref{AppBI}.}.  This method will be used in the next section to find the Radial-Gram-Charlier distribution for arbitrary harmonics.

The characteristic function for a one-dimensional distribution is $\la e^{i k x}\ra$, and the cumulants $\kappa_n$ of such a distribution is found from  $\log\la e^{i k x}\ra=\sum_{n=1} (ik)^n \kappa_n/n!$. The first few cumulants are presented in the following,
\begin{subequations}\label{1DMomentCumul}
	\begin{eqnarray}
	\kappa_1&=&\la x \ra,\label{1DMomentCumulA}\\
	\kappa_2&=&\la x^2 \ra-\la x \ra^2,\label{1DMomentCumulB}\\
	\kappa_3&=&\la x^3 \ra-3\la x \ra \la x^2 \ra+2\la x \ra^3,\label{1DMomentCumulC}\\
	\kappa_4&=&\la x^4 \ra-4\la x \ra\la x^3 \ra-3\la x^2 \ra^2,\label{1DMomentCumulD}\\
	&&\hspace*{2cm}+12 \la x \ra^2\la x^2 \ra-6\la x \ra^4,\nonumber
	\end{eqnarray}
\end{subequations}
where the averages are performed with respect to $p(x)$ in the right-hand side.

Now, consider an approximation for the original distribution where its cumulants are coincident with the original $p(x)$ only for a few first cumulants. We show this approximated distribution by $p_q(x)$ where the cumulants $\kappa_n$ for $1\leq n \leq q$ are the same as the cumulants of the original $p(x)$. 

Assume the following ansatz for this approximated distribution,
\bea\label{A14}
p_q(x)=\frac{1}{\sqrt{2\pi}\sigma}e^{-\frac{x^2}{2\sigma^2}}\sum_{i=0}^{q}T_i(x),
\eea
where
\bea\label{A15}
T_i(x)=\sum_{k=0}^{i} a_{i,k} x^k \qquad a_{0,0}=1.
\eea
In the above, $a_{i,k}$ (except $a_{0,0}$) are unknown coefficients. Note that $p_0(x)$ is nothing but a Gaussian distribution located at the origin. One can find the unknown coefficients $a_{i,k}$ by using equations in \eqref{1DMomentCumul} together with the normalization condition iteratively. In what follows we show how it works: let us present the moments obtained from $p_q(x)$ as $\la x^m \ra_q$. Also, assume that the first moment (which is the first cumulant  too) is zero. At the end, we recover the first moment by applying a simple shift. Now for the first iteration  ($q=1$) we have $\la 1 \ra_{1}=1+a_{1,0}=1$ from the normalization condition and $\la x \ra_{1}=\sigma^2a_{1,1}=\kappa_1=0$ from the Eq.~\eqref{1DMomentCumulA}. It is a linear two dimensional system of equations and the solution is $a_{1,0}=a_{1,1}=0$. In the next step ($q=2$), we have three equations (one normalization condition and two Eq.~\eqref{1DMomentCumulA} and Eq.~\eqref{1DMomentCumulB}). By considering $\kappa_2=\sigma^2$, we find $a_{2,0}=a_{2,1}=a_{2,2}=0$. However, the third iteration is non-trivial. The equations are
\begin{align*}
\la 1 \ra_{3}&=1+a_{3,0}+\sigma^2a_{3,2}=1,\\
\la x \ra_{3}&=\sigma^2 a_{3,1}+3\sigma^4 a_{3,3}=\kappa_1=0,\\	
\la x^2 \ra_{3}&=\sigma^2+\sigma^2 a_{3,0}+3\sigma^4 a_{3,2}=\sigma^2,\\
\la x^3 \ra_{3}&=3\sigma^4 a_{3,1}+15\sigma^6 a_{3,3}=\kappa_3.\\
\end{align*} 
The above equations can be solved easily, $a_{3,0}=a_{3,2}=0$ and $a_{3,1}=-3\sigma^2 a_{3,3}=-\kappa_3/(2\sigma^4)$ which leads to $T_3(x)=\kappa_3/(6\sigma^3) He_3(x/\sigma)$. Here, $He_n(x)$ is the probabilistic Hermite polynomial defined as $He_n(x)=e^{x^2/2}(-d/dx)^n e^{-x^2/2}$. We are able to continue the iterative calculations to any order and find
\begin{equation}\label{GCA}
\begin{aligned}
p(x)=\frac{1}{\sqrt{2\pi}\sigma} &e^{-\frac{(x-\kappa_1)^2}{2\sigma^2}}\sum_{n=0} \frac{h_n}{n!} He_n((x-\kappa_1)/\kappa_2^{1/2}).
\end{aligned}
\end{equation}
In the above, $h_0=1$ and $h_1=0$ together with
\begin{equation}\label{A5}
\begin{aligned}
h_3&=\gamma_1,\\
h_4&=\gamma_2,\\
h_5&=\gamma_3,\\
h_6&=\gamma_4+10\gamma_1^2,\\
&\;\;\vdots
\end{aligned}
\end{equation}
where $\gamma_n$ are the \textit{standardized cumulants} defined as
\bea\label{1Dstandard}
\gamma_{n-2}=\frac{\kappa_n}{\kappa_2^{n/2}}.
\eea
Note that in the Eq.~\eqref{GCA}, we arbitrarily shifted the distribution to the case that the first moment of $p(x)$ is $\kappa_1$. In addition, we assumed that the width of the Gaussian distribution is exactly equal to $\kappa_2$. The Eq.~\eqref{GCA} is the well-know Gram-Charlier A series for the distribution $p(x)$.

One could consider the Eq.~\eqref{GCA} as an expansion in terms of Hermite polynomials. Using the fact that $He_n(x)$ are orthogonal with respect to the weight $w(x)=e^{-x^2/2}$ $/\sqrt{2\pi}$,
\[ \int_{-\infty}^{\infty}  dx\, w(x) He_m(x)He_n(x) = m! \,\delta_{mn}, \]
we can find the coefficients $h_n$ in Eq.~\eqref{A5} (see Red.~\cite{kendallBook}). To this end, we change the coordinate as $x\to (x-\kappa_1)/\sigma$. As a result, we have $$h_n=\int_{-\infty}^{\infty} dx \; p(x) He_n((x-\kappa_1)/\sigma).$$ By using the series form of the Hermite polynomial, we find $h_n$ as a function of $p(x)$ moments. Rewriting moments in terms of cumulants (reverting the equations in  \eqref{1DMomentCumul}), one finds Eq.~\eqref{GCA}.

\subsection{Radial-Gram-Charlier Distribution}\label{arbitraryHarm}

Using standard methods, we can extend one dimensional Gram-Charlier A series \eqref{GCA} to two dimensions (see Appendix~\ref{AppBII}),
\bea\label{expansion}
\hspace*{-0.6cm}p(\br)=\N(\br)\sum_{\substack{m,n=0}}\frac{h_{mn}}{m!n!} He_m(\frac{x-\mu_x}{\sigma_x})He_n(\frac{y-\mu_y}{\sigma_y}),
\eea
where $h_{mn}$ are written in terms of two dimensional cumulants $\A_{mn}$ (see Eqs.~\eqref{hmnTwo}-\eqref{hmnSix}), and $\N(\br)$ is a two dimensional Gaussian distribution similar to Eq.~\eqref{2DGauss} located at $(\mu_x,\mu_y)$ and $\sigma_x\neq\sigma_y$. 

It is worth noting that the concept of 2D Gram-Charlier A series has been employed in heavy ion physics first in Ref.~\cite{Teaney:2010vd} by Teaney and Yan. They used this series to study the energy density of a single event\footnote{We explicitly connect the Eq.~\eqref{expansion} to the results in Ref.~\cite{Teaney:2010vd} in the Appendix~\ref{AppBIII}.}.  However, we use this to study flow harmonic distribution in the present work. 

Now let us consider a two dimensional Gram-Charlier A series for $p(v_{n,x},v_{n,y})$. By this consideration, one can find a corresponding series for $p(v_n)$ by averaging out the azimuthal direction. We should say that the result of this averaging for the second and third harmonics are different. For $n=3$,  the distribution $p(v_{3,x},v_{3,y})$ is rotationally symmetric and, as we already remarked in the previous section, the whole information of the distribution is encoded in $c_3\{2k\}$. As a result, we are able to rewrite the 2D cumulants $\A_{mn}$ in terms of $c_3\{2k\}$. It has been done in Ref.~\cite{Abbasi:2017ajp}, and an expansion for $p(v_3)$ has been found. On the other hand, for $n=2$ the whole information of $p(v_{2,x},v_{2,y})$ is not in $c_2\{2k\}$. Therefore, we are not able to rewrite all $\A_{mn}$ in terms of $c_2\{2k\}$ after averaging out the azimuthal direction of a 2D Gram-Charlier A series.

For completeness, we studied the azimuthal averaging of Eq.~\eqref{expansion} in the most general case in the Appendix~\ref{AppC}. In this appendix, we showed that the distribution in Ref.~\cite{Abbasi:2017ajp} is reproduced only by assuming $\A_{10}=\A_{01}=0$. Also, we discussed about the information we can find from the averaged distribution compared to the two dimensional one. However, the method which we will follow in this section is different from that point out in Appendix~\ref{AppC}. Consequently, the most general series we will find here is not coincident with the distribution obtained in the Appendix~\ref{AppC}.

Before finding a Gram-Charlier A series for arbitrary harmonic, let us find the series for odd harmonics (mentioned in Ref.~\cite{Abbasi:2017ajp}) by employing orthogonal polynomials. The result will be used to find the series for the most general case later. Since we have $\vb_3=0$ for $n=3$, the Bessel-Gaussian distribution reduces to a radial Gaussian distribution as $(v_3/\sigma^2) e^{-v_3^2/(2\sigma^2)}$. Moreover, the Laguerre polynomials $L_n(x)$ are orthogonal with respect to the weight $w(x)=e^{-x}$ in the range $[0,\infty)$,
$$ \int_{0}^{\infty} dx\,e^{-x} L_n(x)L_m(x)=\delta_{mn}.$$
By changing the coordinate as $x=v_3^2/(2\sigma^2)$, the measure $w(x)dx$ changes to $(v_3 /\sigma^2) e^{-v_3^2/(2\sigma^2)} dv_3$; the radial Gaussian distribution appears as the weight for orthogonality of $L_n(v_3^2/(2\sigma^2))$. Hence, we can write a general distribution $p(v_3)$ like 
\bea\label{RGCODD}
p(v_3)=\frac{v_3}{\sigma^2} e^{-\frac{v_3^2}{2\sigma^2}}\sum_{n=0}^{\infty} \frac{(-1)^n\ell^{\text{odd}}_{2n}}{n!} L_n(v_3^2/(2\sigma^2)),
\eea
where the coefficients $\ell^{\text{odd}}_{2n}$ can be found by\footnote{In Eq.~\eqref{RGCODD}, we chose the coefficient expansion as  $\frac{(-1)^n\ell^{\text{odd}}_{2n}}{n!}$  for convenience. }
\bea
\ell^{\text{odd}}_{2n}=n!(-1)^n\int_{0}^{\infty} dv_3 \; p(v_3) L_n(v_3^2/(2\sigma^2)). 
\eea
Considering the series form of the Laguerre polynomial,
\bea\label{LaguExpand}
L_n(x)=\sum_{k=0}^{n}\binom{n}{k} \frac{(-1)^k}{k!} x^k,
\eea
we immediately find $\ell^{\text{odd}}_n$ in terms of moments $\la v_3^{2q} \ra $. Then one can invert the equations in \eqref{1DnPartCumul} to write the moments $\la v_3^{2q} \ra$ in terms of cumulants $c_3\{2q\}$. If we do so and by choosing $2\sigma^2=\la v_3^2 \ra=c_3\{2\}$ \footnote{Refer to the footnote \ref{footnote} and set $\A_{10}=0$.}, we find $\ell^{\text{odd}}_0=1$ and $\ell^{\text{odd}}_2=0$ together with
\begin{equation}\label{laguerreQs}
\begin{aligned}
\ell_4^{\text{odd}}&=\Gamma_2^{\text{odd}},\\
\ell_6^{\text{odd}}&=\Gamma_4^{\text{odd}},\\
\ell_8^{\text{odd}}&=\Gamma_6^{\text{odd}}+18(\Gamma_2^{\text{odd}})^2,\\
\ell_{10}^{\text{odd}}&=\Gamma_8^{\text{odd}}+100(\Gamma_2^{\text{odd}})(\Gamma_4^{\text{odd}}),
\end{aligned}
\end{equation}
where in the above we defined the standardized cumulants $\Gamma^{\text{odd}}_{2k}$ as
\bea\label{gamma0}
\Gamma_{2k-2}^{\text{odd}}=\frac{c_3\{2k\}}{c_2^k\{2\}},
\eea
similar to Eq.~\eqref{1Dstandard}.

The expansion \eqref{RGCODD} together with Eq.~\eqref{laguerreQs} is exactly the series found in Ref.~\cite{Abbasi:2017ajp} which is true for any odd $n$. This approximated distribution is called the Radial-Gram-Charlier (RGC) distribution in Ref.~\cite{Abbasi:2017ajp}.

It should be noted that it is a series for the case that $\vb_n=0$. 
In the following, we will try to find similar series for $p(v_n;\vb_n)$ where $\vb_n$ could be non-vanishing.

In order to find the Gram-Charlier A series for the distribution $p(v_n,;\vb_n)$ in general case, we comeback to the iterative method explained in Sec.~\ref{IIIA} where we found the distribution \eqref{GCA} by considering the ansatz \eqref{A14} and iteratively solving the equations in \eqref{1DMomentCumul}. 

Here, we assume that $p_q(v_n;\vb_n)$ is an approximation of $p(v_n)$ where only the cumulants $c_n\{2k\}$ with $1\leq k\leq q$ are same as the original distribution. Now suppose an ansatz that has the following properties:
\begin{itemize}
	\item Its leading order corresponds to the Bessel-Gaussian distribution.
	\item In the limit $\vb_n\to 0$, the distribution approaches to \eqref{RGCODD}.
\end{itemize}
Using such an ansatz, we calculate the moments $\la v_n^{2k} \ra$ with some unknown parameters and  find them by solving the equations in \eqref{1DnPartCumul} iteratively. 

We introduce the following form for the ansatz,
\bea\label{GGCD}
p_q(v_n;\vb_n)=\left(\frac{v_n}{\sigma^2}\right)  e^{-\frac{v_n^2+\vb_n^2}{2\sigma^2}}\sum_{i=0}^{q}\Q_i(v_n;\vb_n),
\eea
where
\bea\label{QDist}
\Q_i(v_n;\vb_n)=\sum_{k=0}^{i} a_{i,k} \left(-\frac{v_n}{\vb_n}\right)^k\,I_k\left(\frac{v_2 \vb_2}{\sigma^2}\right).
\eea
One simply finds that by choosing $a_{0,0}=1$ the distribution $p_0(v_n;\vb_n)$ is the Bessel-Gaussian distribution. On the other hand, the function $\Q_i(v_n;\vb_n)$ in the limit $\vb_n \to 0$ reduces to
\bea
\sum_{k=0}^{i} a_{i,k} \frac{(-1)^k}{k!} \left(\frac{v_n}{2\sigma^2}\right)^k.
\eea
By comparing the above expansion with Eq.~\eqref{LaguExpand},  we realize that if we choose $a_{i,k}\propto \binom{i}{k}$, $\Q_i(v_n;\vb_n)$ is proportional to $L_n(v_n^2/(2\sigma^2))$. In order to reproduce the Eq.~\eqref{RGCODD}, we choose
\bea\label{ansatz3}
a_{i,k} = \ell_{2i} \,\frac{(-1)^i}{i!}\,\binom{i}{k},
\eea
where $\ell_{2i}$ are unknown coefficients.

Similar to the Sec.~\ref{IIIA}, we indicate the moments of $p_q(v_n;\vb_n)$ by $\la v_n^{2m} \ra_q$. For the first iteration, we have the normalization condition $\la 1 \ra_0=\ell_0=1$. For the second iteration, the normalization condition is trivially satisfied $\la 1 \ra_1=1$ while from the Eq.~\eqref{1DnPartCumulA} we have $\la v_n^2 \ra_1 = \vb_n^2+2\sigma^2 (1+\ell_2)=c_n\{2\}$. At this stage, we choose $\ell_2=0$ to have\footnote{ \label{footnote}  Obviously, there is no one-to-one correspondence between $c_n\{2k\}$ and $\A_{mn}$ due to the losing information by averaging. Specifically, one can find $c_n\{2\}=\A_{10}^2+\A_{01}^2+\A_{20}+\A_{02}$ (see Eq.~\eqref{cn2ToAmnA}).
	Note that by assuming $\Phi_{\text{RP}}=0$ we have $\A_{01}=\la v_{n,y}\ra =0$ and $\A_{10}=\la v_{n,x}\ra =\vb_n$. Also, it is a reasonable assumption to consider that $\A_{20}\simeq \A_{02}$ (see  Refs.~\cite{Giacalone:2016eyu,Abbasi:2017ajp}). By choosing $\sigma=\sigma_x=\sigma_y=\A_{20}\simeq \A_{02}$, one approximates $p(v_{n,x},v_{n,y})$ around a symmetric Gaussian distribution located at $(\vb_n,0)$ where its width is exactly similar to the distribution $p(v_{n,x},v_{n,y})$. In this case, we find $c_n\{2\}=\vb_n^2+2\sigma^2$. }
\bea\label{cn2Approx}
c_n\{2\}=\vb_n^2+2\sigma^2.
\eea

For the next iteration, we find that the normalization condition  and Eq.~\eqref{1DnPartCumulA} are automatically satisfied, $\la 1 \ra_2=1$, $\la v_n^2 \ra_2=c_n\{2\}$ while the Eq.~\eqref{1DnPartCumulB} leads to a non-trivial equation for $\ell_4$,
\bea
c_n\{6\}=\ell_4(c_n\{2\}-\vb_n^2)^2-\vb_n^4.
\eea
Using the above equation, we immediately find $\ell_4$. In a similar way, we are able to continue this iterative calculation and find $\ell_{2q}$ from the only non-trivial algebraic equation at each step. A summary of the few first results are as follows: $\ell_0=1$ and $\ell_2=0$ together with
\begin{equation}\label{generL}
\begin{aligned}
\ell_4&=\frac{c_n\{4\}+\vb_n^4}{(c_n\{2\}-\vb^2)^2},\\
\ell_6&=\frac{c_n\{6\}+6 c_n\{4\}\vb_n^2+2\vb_n^6}{(c_n\{2\}-\vb^2)^3},\\
\ell_8&=\frac{c_n\{8\}+12 c_n\{6\} \vb_n^2+18 c_n^2\{4\}+42 c_n\{4\} \vb_n^4+9\vb_n^8}{(c_n\{2\}-\vb^2)^4}.
\end{aligned}
\end{equation}

One may wonder that, similar to two previous cases, are we able to write all the coefficients $\ell_{2k}$ in terms of some standardized cumulants? These standardized cumulants should smoothly approach to what is mentioned in Eq.~\eqref{gamma0}. In fact, it motivates us to define a new set of cumulants,
\begin{subequations}\label{newCumuls}
	\begin{eqnarray}
	q_n\{2\}&=&c_n\{2\}-\vb_n^2,\label{newCumulsA}\\
	q_n\{4\}&=&c_n\{4\}+\vb_n^4,\label{newCumulsB}\\
	q_n\{6\}&=&c_n\{6\}+6 c_n\{4\}\vb_n^2+2\vb_n^6,\label{newCumulsC}\\
	q_n\{8\}&=&c_n\{8\}+12 c_n\{6\} \vb_n^2+6c_n\{4\}\vb_n^4-9\vb_n^8,\quad\label{newCumulsD}\\
	q_n\{10\}&=&c_n\{10\}+20(c_n\{8\}-12 c_n^2\{4\})\vb_n^2,\label{newCumulsE}\\
	&&+30\, c_n\{6\} \vb_n^4-480\,c_n\{4\} \vb_n^6-156\, \vb_n^{10}.\nonumber
	\end{eqnarray}
\end{subequations}
Now, we can define the standardized form of this new sets of cumulants as follows,
\bea\label{gammaDefi}
\Gamma_{2k-2}=\frac{q_n\{2k\}}{q_n^k\{2\}}.
\eea
Using the above definitions, we can rewrite the coefficients $\ell_{2k}$ in the following form,
\begin{equation}\label{genL2}
\begin{aligned}
\ell_4&=\Gamma_2,\\
\ell_6&=\Gamma_4,\\
\ell_8&=\Gamma_6+18\Gamma_2^2,\\
\ell_{10}&=\Gamma_8+100\Gamma_2\Gamma_4,
\end{aligned}
\end{equation}
which are in agreement with equations in \eqref{laguerreQs} in the limit $\vb_n\to 0$.

Let us to summarize the series in \eqref{GGCD} as follows
\bea\label{GGCD2}
p(v_n;\vb_n)=\frac{v_n}{\sigma^2}  e^{-\frac{v_n^2+\vb_n^2}{2\sigma^2}}\sum_{i=0}^{\infty}\frac{(-1)^i\ell_{2i}}{i!} \tilde{\Q}_i(v_n;\vb_n),\hspace*{0.6cm}
\eea
where $\tilde{\Q}_i(v_n;\vb_n)$ is similar to $\Q_i(v_n;\vb_n)$ in Eq.~\eqref{QDist} up to a numerical factor,
\begin{equation}\label{tildeQ}
\begin{aligned}
\tilde{\Q}_i(v_n;\vb_n)&=\frac{(-1)^i \ell_{2i}}{i!} \Q_i(v_n;\vb_n)\\
&=\sum_{k=0}^{i}\binom{i}{k}\left(-\frac{v_n}{\vb_n}\right)^k\,I_k\left(\frac{v_n \vb_n}{\sigma^2}\right).
\end{aligned}
\end{equation}

Recall that both  distributions \eqref{GCA} and \eqref{RGCODD} could be found by using the orthogonality of $He_n(x)$ and $L_n(x)$. We can ask that is there any similar approach to find Eq.~\eqref{GGCD2}? Surprisingly, $\tilde{\Q}_i(v_n;\vb_n)$ is related to a generalized class of orthogonal polynomials which are called \textit{multiple orthogonal polynomials} (see Ref.~\cite{ismail}). These generalized version of the polynomials are orthogonal with respect to more than one weight. Specifically, the polynomials related to $\tilde{\Q}_i(v_n;\vb_n)$ have been introduced in Ref.~\cite{CoussementVanAssche}.  In order to avoid relatively formal mathematical material here, we refer the interested reader to Appendix~\ref{appD} where we briefly review the multiple orthogonal polynomials and  re-derive Eq.~\eqref{GGCD2} by employing them.

\begin{figure}[t!]
	\begin{center}
		\begin{tabular}{c}
			\includegraphics[scale=0.4]{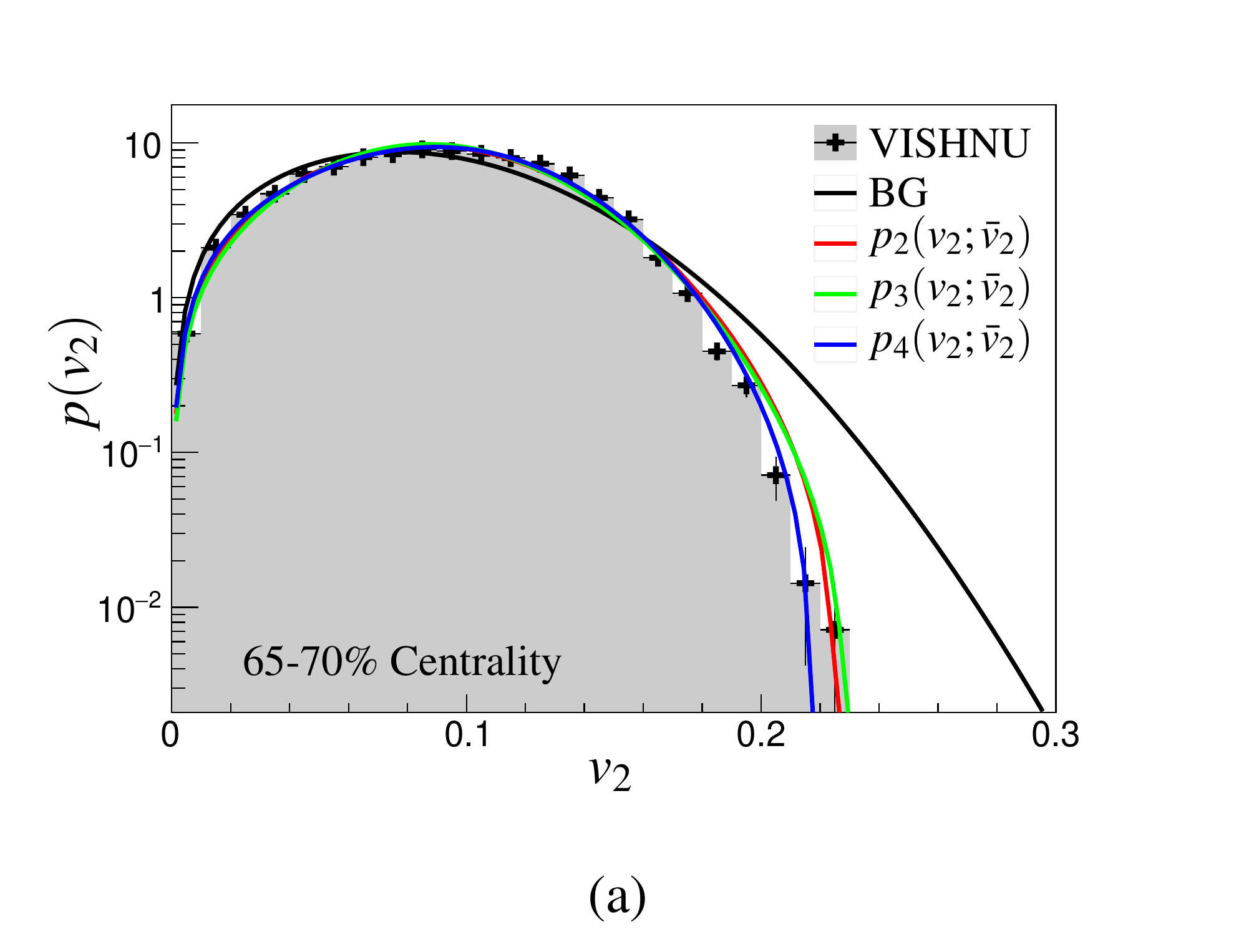}\\
			\includegraphics[scale=0.4]{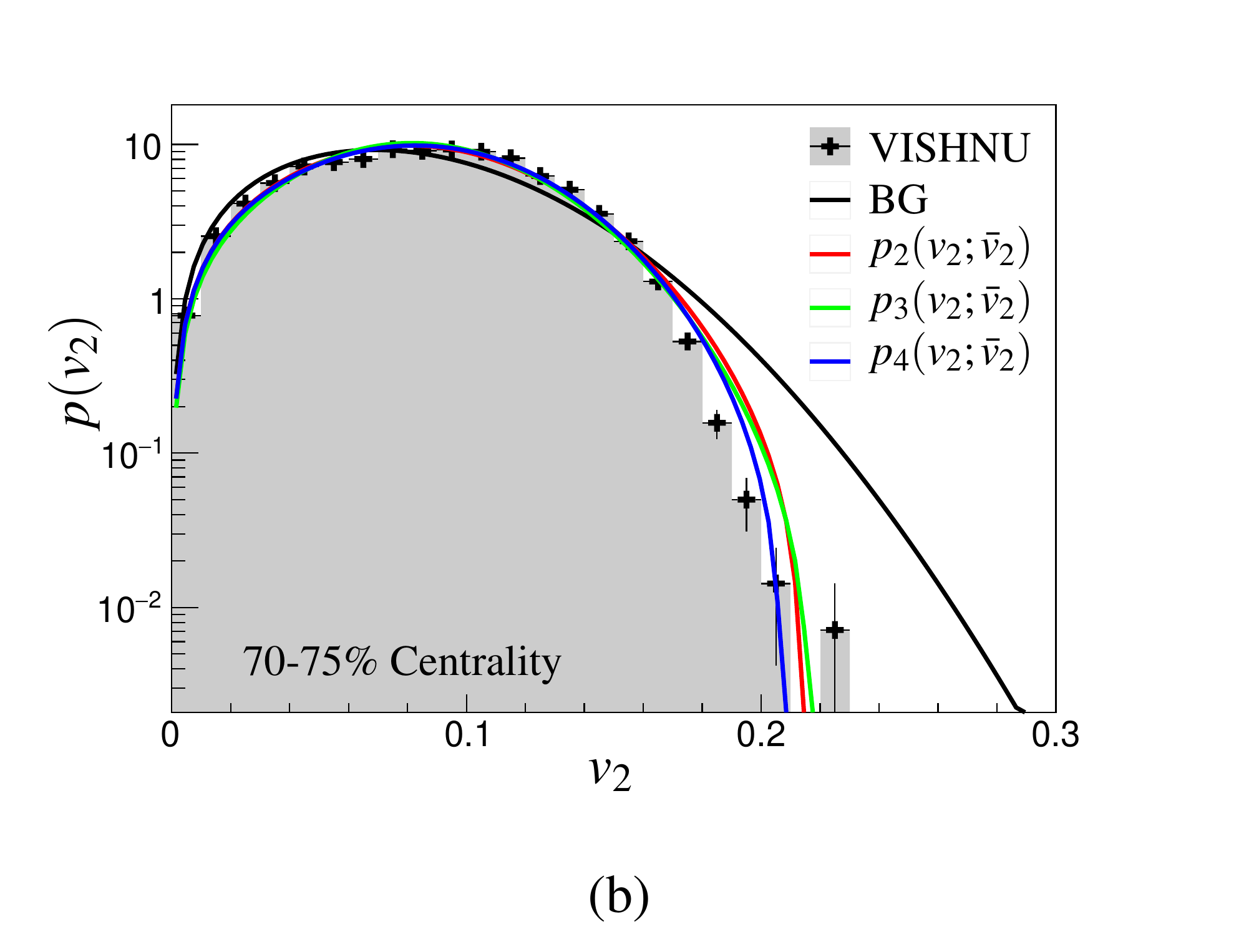}\\
			\includegraphics[scale=0.4]{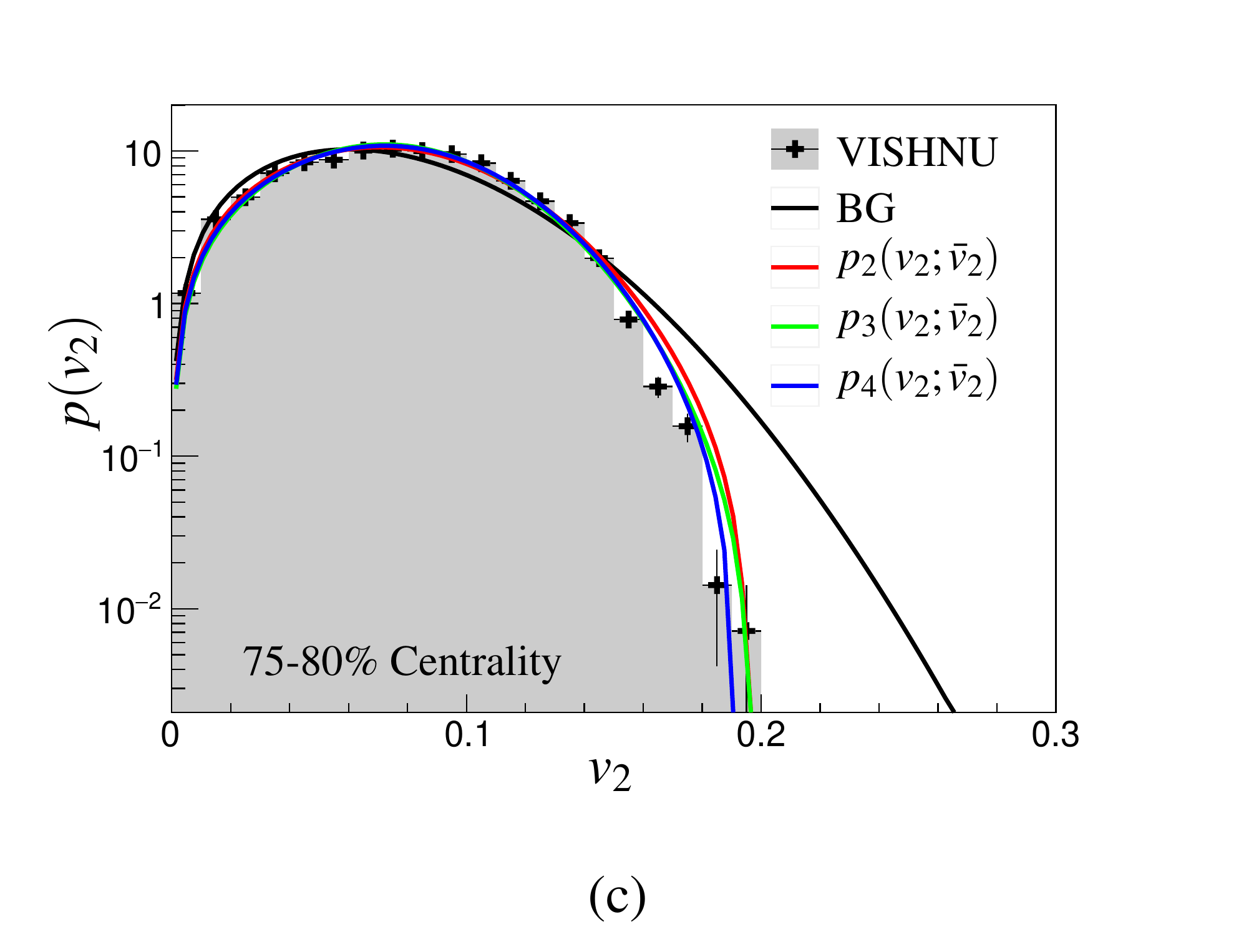} 			 			
		\end{tabular}		
		\caption{Comparing the flow harmonic distribution from iEBE-VISHNU (shaded region) and Gram-Charler A series different approximations.} 
		\label{ConverganceCheck}
	\end{center}
\end{figure}

\begin{figure}[t!]
	\begin{center}
		\begin{tabular}{c}
			\includegraphics[scale=0.4]{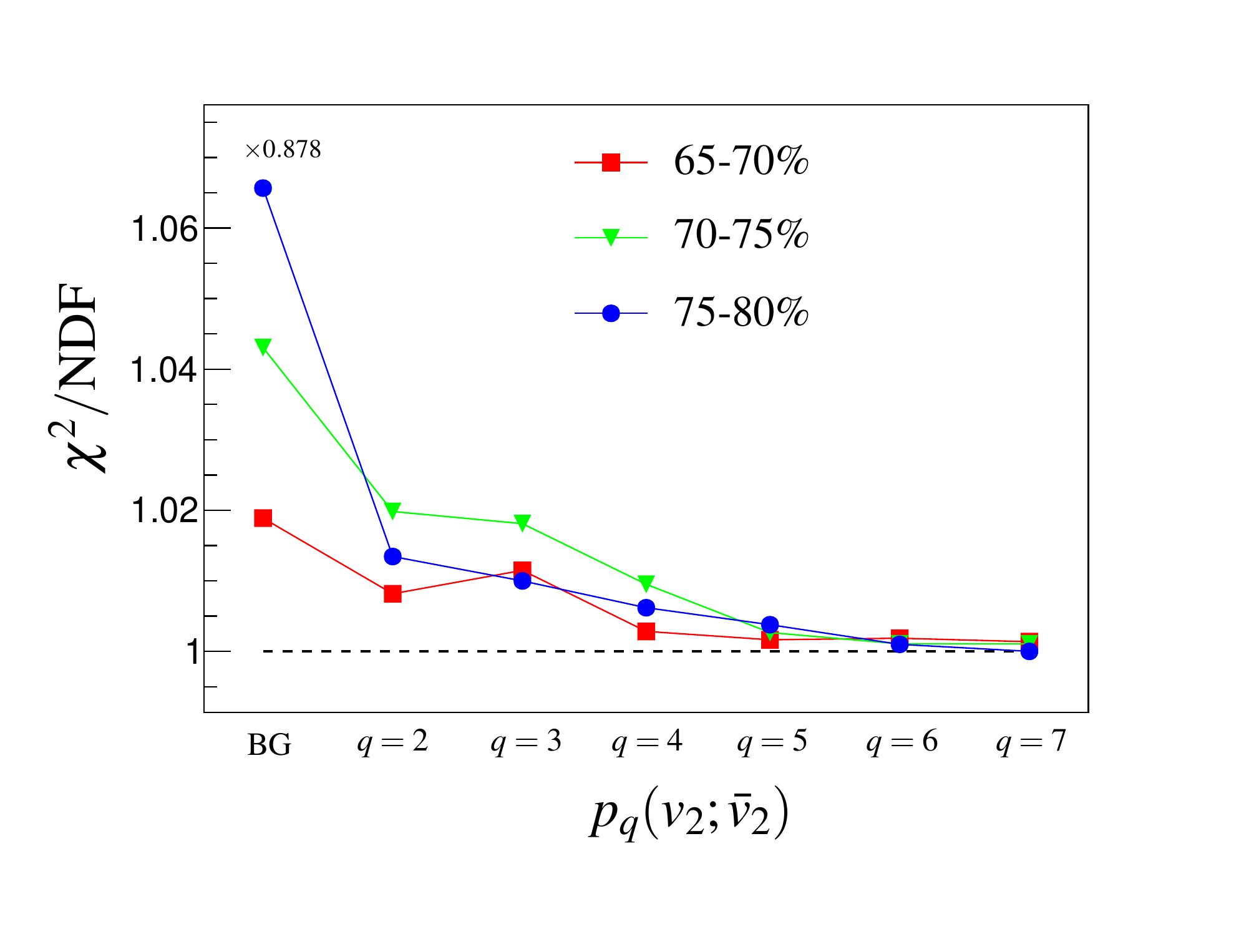} 			 			
		\end{tabular}		
		\caption{Examining the accuracy of the Gram-Charlier A series with the actual distribution obtained from simulation by studying $\chi^2/\text{NDF}$.} 
		\label{ConverganceCheckChi2}
	\end{center}
\end{figure}

An important point about the distribution \eqref{GGCD2} is the convergence of its summation. For sure, finding the convergence condition of the infinite sum in Eq.~\eqref{GGCD2} is beyond the scope of the present paper. However, if we find that at least a few first terms in Eq.~\eqref{GGCD2} is sufficient to give a reasonable approximation of $p(v_n)$ then there is no concern about the convergence or divergence of this series practically\footnote{This is an argument presented in Ref.~\cite{HCramer}. In the same reference the convergence condition of Eq.~\eqref{GCA} (which is not our main interest here) can be found. It is shown that if $p(x)$ is a function of bounded variation in the range $(-\infty,\infty)$ and the integral $\int_{-\infty}^{\infty} dx\,e^{-x^2/4}p(x)$ is convergent then the series \eqref{GCA} is convergent. Otherwise it might diverge.}.

In order to show how much the distribution \eqref{GGCD2} is a good approximation, we need to have a sample for $p(v_n)$ where its $\vb_n$ is known. To this end, we generate heavy ion collision events by employing a hydrodynamic based event generator which is called iEBE-VISHNU \cite{Shen:2014vra}. The reaction plane angle is set to zero in this event generator. Thus, we can simply find $p(v_{n,x}v_{n,y})$ and subsequently $\vb_n$. The events are  divided into sixteen centrality classes between 0 to 80 percent and at each centrality class we generate 14000 events. The initial condition model is set to be MC-Glauber.

Let us recover the notation in Eq.~\eqref{GGCD} and assume that $p_q(v_n;\vb_n)$ is the distribution \eqref{GGCD2} where the summation is done up to $i=q$.  We first compute the $c_2\{2k\}$ and $\vb_2$ from iEBE-VISHNU output and plug the results in Eq.~ \eqref{GGCD2}. After that we can compare the original simulated distribution $p(v_n)$ with estimated $p_q(v_2;\vb_2)$. The results are presented in the Fig.~\ref{ConverganceCheck} for the events in 65-70$\%$, 70-75$\%$ and 75-80$\%$ centrality classes in which we expect the distribution is deviated from the Bessel-Gaussian. In this figure, the black curve corresponds to the Bessel-Gaussian distribution ($p_0(v_n;\vb_n)$) and the red, green and blue curves correspond to $p_q(v_2;\vb_2)$ with $q=2$, $3$ and $4$, respectively.  Recall that  $q=1$ has no contribution because $\ell_2$ vanishes. As can be seen in the figure, the black curve shows that the distribution is deviated from Bessel-Gaussian and distribution $p_q(v_2;\vb_2)$ with $q\neq0$ explains the generated data more accurately.

\begin{figure*}[th!]
	\begin{center}
		\begin{tabular}{cc}
			\hspace*{-0.8cm}	\includegraphics[scale=0.43]{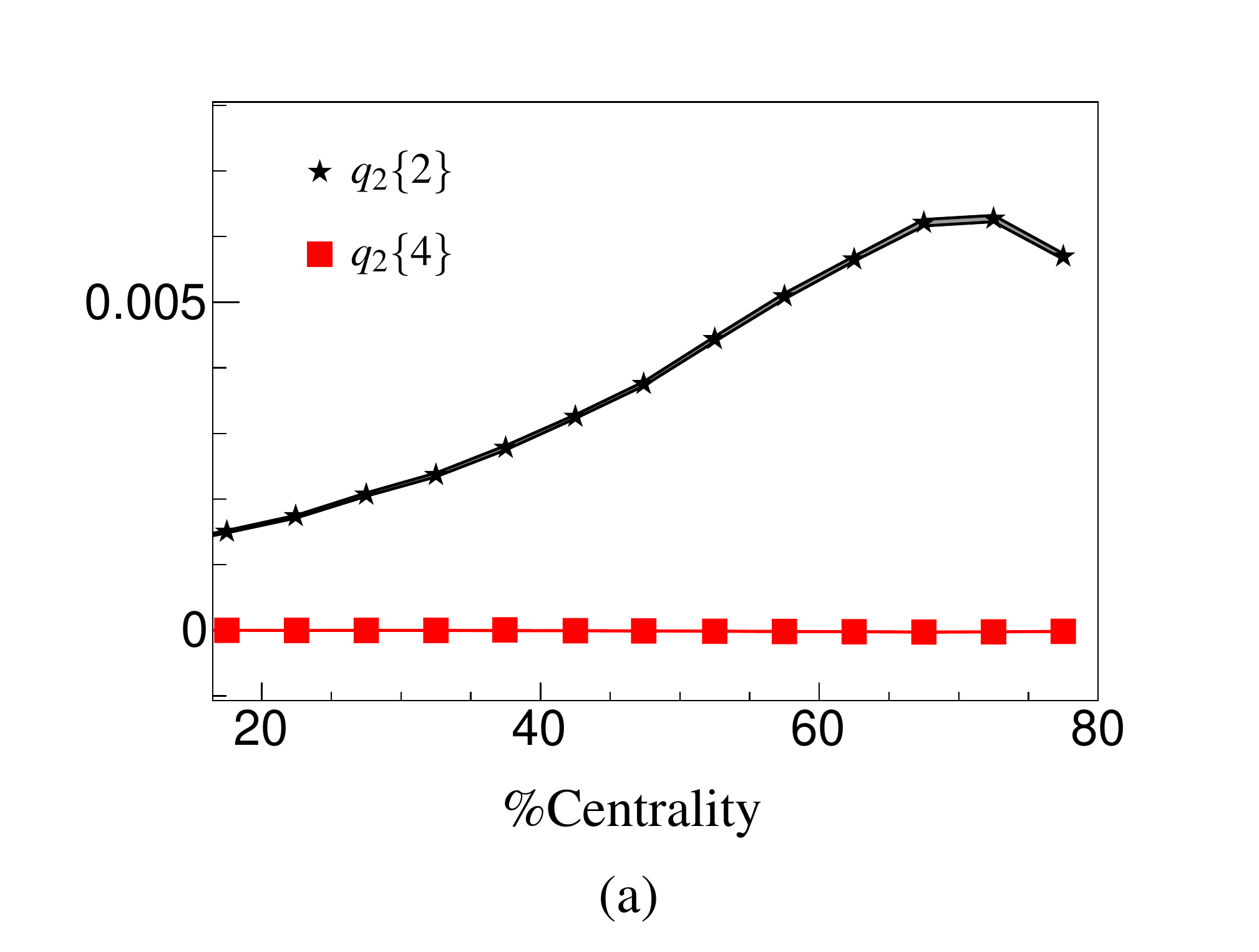} &
			\includegraphics[scale=0.43]{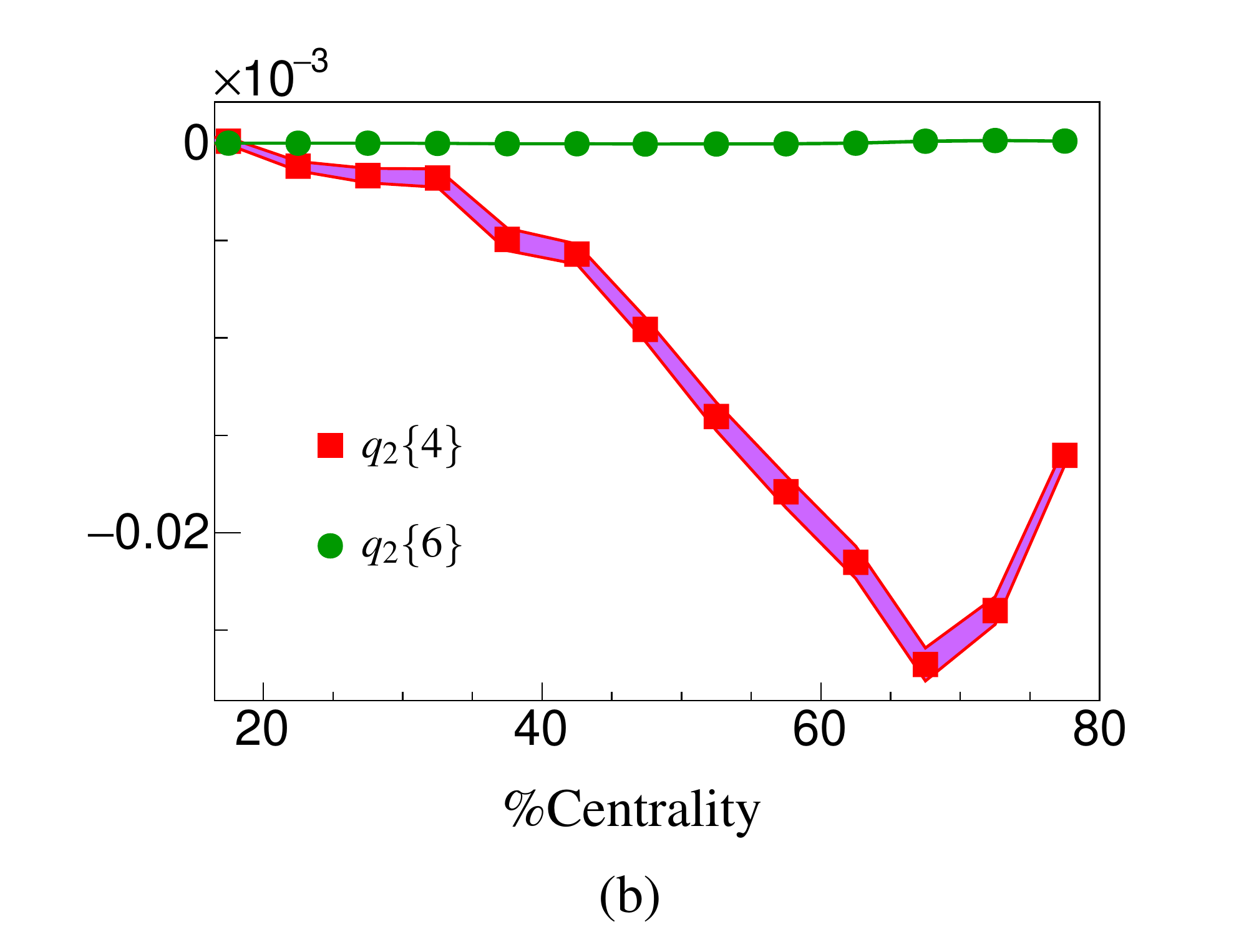} \\
			\hspace*{-0.8cm}	\includegraphics[scale=0.43]{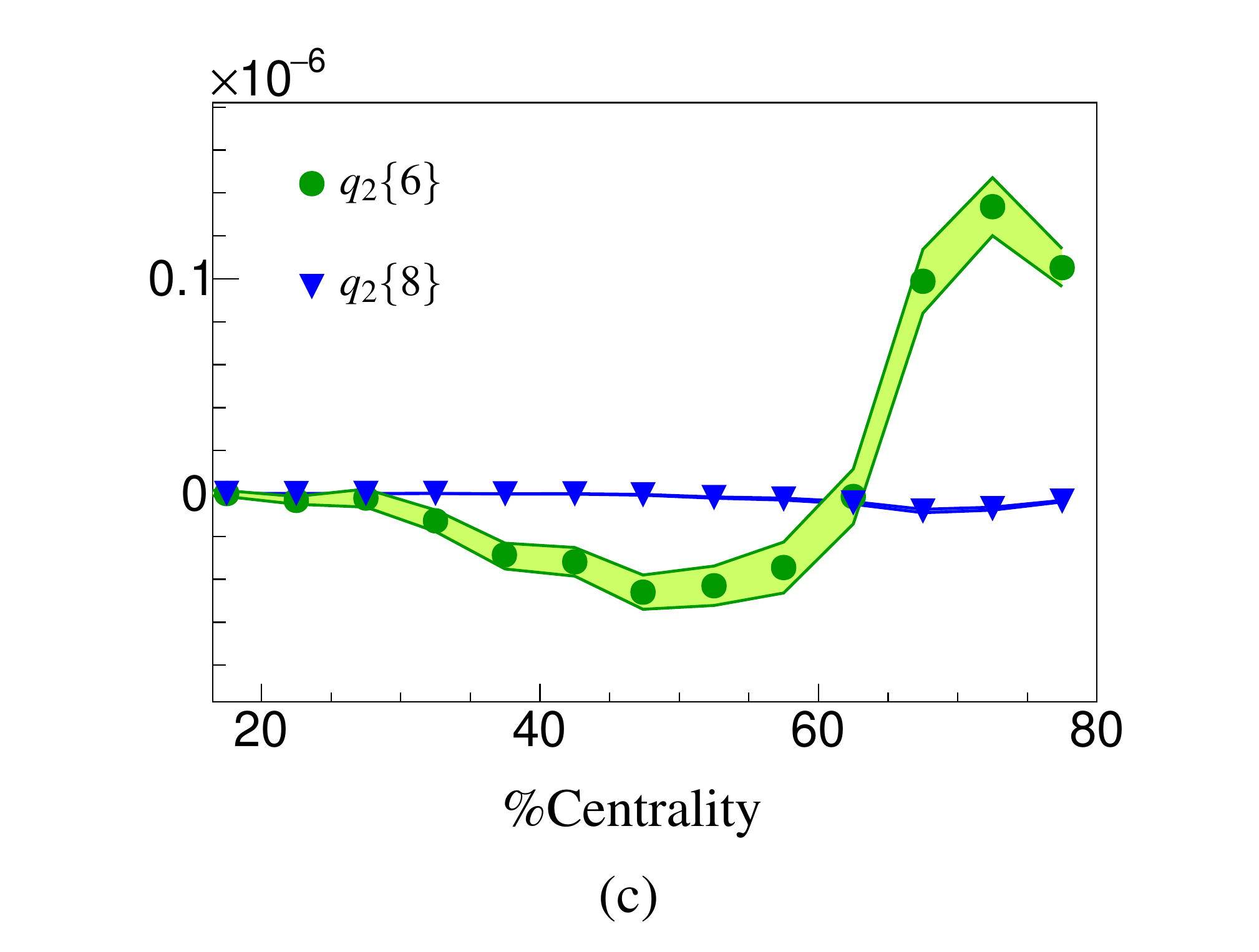} &
			\includegraphics[scale=0.43]{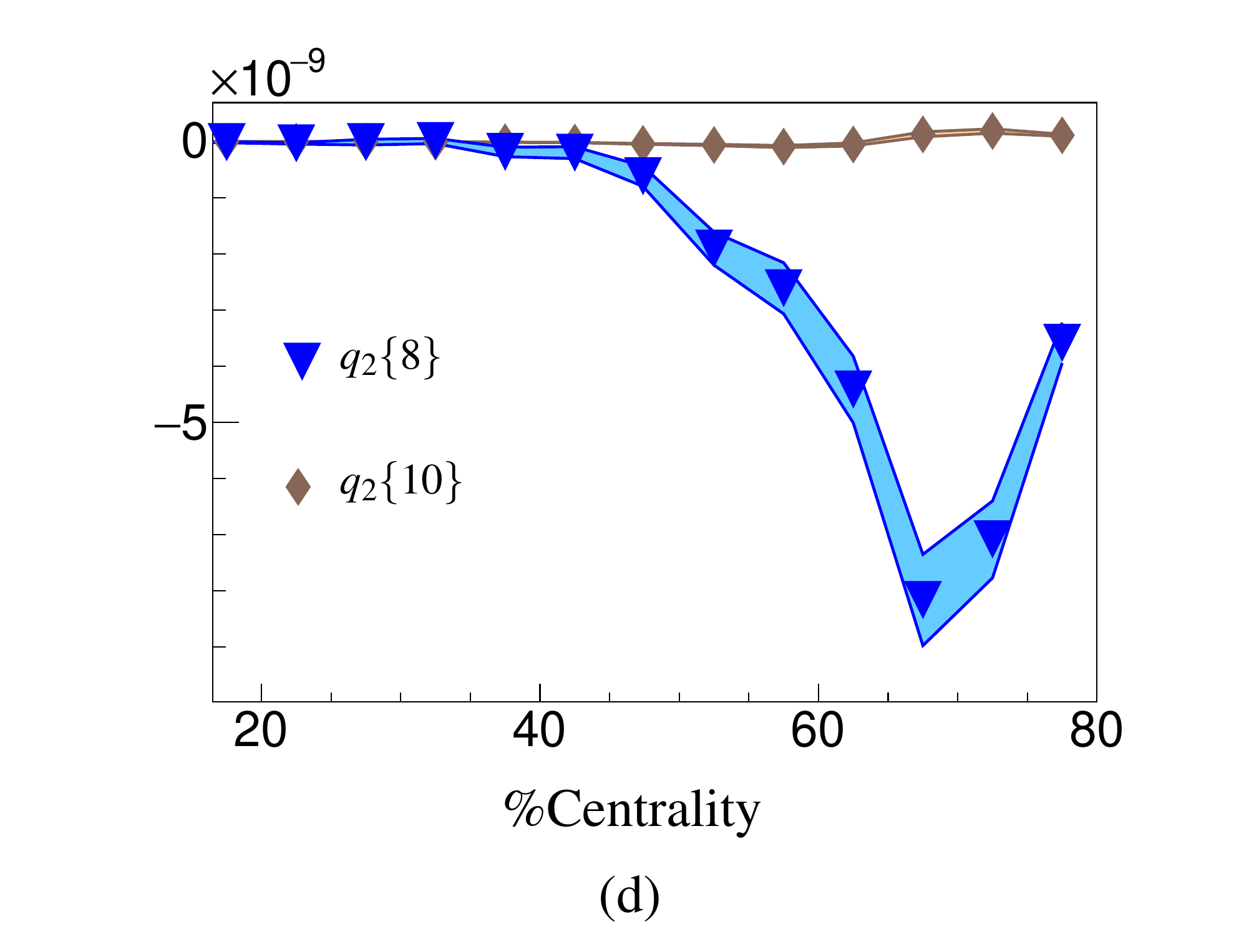} 			
		\end{tabular}		
		\caption{The cumulants $q_2\{2k\}$ obtained from the iEBE-VISHNU event generator. } 
		\label{qmqn}
	\end{center}
\end{figure*}

In order to compare the estimated distributions more quantitatively, we plotted $\chi^2/\text{NDF}$ which compare the estimated distribution $p_q(v_2;\vb_2)$ with iEBE-VISHNU output. We plotted the results in Fig.~\ref{ConverganceCheckChi2} for $q=0,2,3,4,5,6$ and $7$ for the events in 65-70$\%$, 70-75$\%$ and 75-80$\%$ centrality classes. The value of $\chi^2/\text{NDF}$ associated with the Bessel-Gaussian distribution is much greater than others. Therefore, we multiplied its value by $0.878$ to increase the readability of the figure.

As the Fig.~\ref{ConverganceCheckChi2} demonstrates, the Bessel-Gaussian distribution has less compatibility with the distribution. In addition, the quantity $\chi^2/\text{NDF}$ becomes more close to one by increasing $q$\footnote{As an exception, the quantity $\chi^2/\text{NDF}$ increases slightly in moving from $q=2$ to $q=3$ for the events in the $65$-$70\%$ centrality class. However, the overall trend is decreasing in general.}. It is relatively close to one for higher values of $q$. One may deduce from the figure that the series converges because $\chi^2/\text{NDF}$ for $q=6$ and $q=7$ are very close to each other and very close to one. However, we should say that although it is an strong evidence for the series convergence, there is no guarantee that by adding higher terms the series remain stable. Additionally, we checked the convergence of the series for the case interested in heavy ion physics. This convergence might not be true for an arbitrary distribution in general.  In any case, what we learn from the above arguments is that at least few first terms in the series \eqref{GGCD2} gives a good approximation compared to the original flow harmonic distributions.

\subsection{New Cumulants}

Let us come back to the cumulants in Eq.~\eqref{newCumuls} and point out their properties. Considering the \textit{new cumulants} $q_n\{2k\}$, we note the following remarks:
\begin{itemize}
	\item Referring to Eq.~\eqref{deltaVn}, all the cumulants $q_n\{2k\}$ for $k\geq 1$ are vanishing for the distribution $\delta(v_{n,x}-\vb_n,v_{n,y})$.
	\item Referring to Eq.~\eqref{BGCUMUls}, the only non-zero $q_n\{2k\}$ for Bessel-Gaussian distribution is $q_n\{2\}=2\sigma^2$.
	\item In the limit $\vb_n\to 0 $, the cumulants $q_n\{2k\} $ approach  to  $c_n\{2k\}$.	
\end{itemize}
The above remarks are indicated that $q_n\{2k\}$ contains information originated from the fluctuations only and the \textit{explicit} effect of the collision geometry $\vb_n$ is extracted from it\footnote{It is important to note that although we extracted the \textit{explicit} collision geometry effect but its footprint still exists in the fluctuations implicitly, e.g. for $\vb_2\neq 0$, the distribution is skewed while for $\vb_2= 0$ is not.  }.   

It is important to note that although we have found $q_n\{2k\}$ by RGC distribution inspiration, we think it is completely independent of that and there must be a more direct way to find $q_n\{2k\}$ independent of the RGC distribution.

Concerning the difference between $c_n\{2k\}$ and $q_n\{2k\}$ in terms of Gram-Charlier expansion, we should say that the cumulants $c_n\{2k\}$ appears as the coefficients of the expansion when we expand the distribution $p(v_n)$ around a radial-Gaussian distribution (see Eq.~\eqref{RGCODD}) while $q_n\{2k\}$ are those appear in the expansion around the Bessel- Gaussian distribution. Now, if the distribution under study is more Bessel-Gaussian rather than the radial-Gaussian, we need infinitely many $c_n\{2k\}$ cumulants to reproduce the correct distribution. For instance, for second harmonics all $v_2\{2k\}$ are non-zero and have approximately close values. It is because we are approximating a distribution which is more Bessel-Gaussian rather than a radial-Gaussian. On the other hand for third harmonics, we expect that the underling distribution is more radial-Gaussian, and practically we see a larger difference between $v_3\{2\}$ and $v_3\{4\}$ compared to the second harmonics \cite{Aad:2014vba}. Based on the above arguments, we deduce that $q_n\{2k\}$ are more natural choice for the case that $\vb_n$ is non-vanishing.

Nevertheless,  the cumulants $q_n\{2k\}$ (unlike $c_n\{2k\}$) are not experimentally observable because of the presence of $\vb_n$ in their definition. However, they are useful to systematically estimate the distribution $p(v_n)$ and consequently estimate the parameter $\vb_n$. This will be the topic of the next section.

\section{Averaged Ellipticity and Flow Harmonic Fine-Splitting}\label{findAverage}

In this section, we would like to exploit the cumulants $q_n\{2k\}$  to find an estimation for $\vb_n$.
Note that if we had a prior knowledge about one of the $q_2\{2k\}$ or even any function of them (for instance $g(q_2\{2\},q_2\{4\},\ldots)$), we could find $\vb_n$  \textit{exactly} by solving the equation $g(q_2\{2\},q_2\{4\},\ldots)$ $=0$ in principle. Because the cumulants $c_n\{2k\}$  are experimentally accessible, one would practically solve an equation $g(\vb_n)=0$. We regret that we have not such a prior knowledge about $q_2\{2k\}$, but we are still able to estimate $\vb_n$ approximately by assuming some properties for $p(v_n)$.

Any given distribution can be quantified by $q_n\{2k\}$.  While $p(v_n)$ is approximately Bessel-Gaussian, we can guess that 
$$q_n\{2\}\gg q_n\{4\} \gg q_n\{6\} \gg q_n\{8\}\gg \cdots\,.$$
In fact, it is confirmed by the simulation. The cumulants $q_2\{2k\}$ are obtained from the iEBE-VISHNU output and presnted in Fig.~\ref{qmqn}. Therefore, as already remarked, we expect that a few first cumulants $q_n\{2k\}$ is enough to quantify the main features of a distribution near Bessel-Gaussian.

Let us concentrate on $n=2$ from now on. Recall that $q_2\{4\}=q_2\{6\}=\cdots=0$ corresponds to Bessel-Gaussian distribution. This choice of cumulants equivalent to a distribution with $v_2\{4\}=v_2\{6\}=\cdots$ which is not compatible with the splitting of $v_2\{2k\}$ observed in the experiment. As we discussed at the beginning of this chapter,  we can find $\vb_2$ by estimating any function of cumulants $q_2\{2k\}$. Here we use the most simple guess for this function which is $g(q_2\{2\},q_2\{4\},\ldots)=q_2\{2k\}$. Therefore, the equation $q_2\{2k\}=0$ for each $k \geq 1$ corresponds to an specific estimation for $p(v_2)$. 

\begin{figure}[t!]
	\begin{center}
		\hspace*{-0.8cm}		\includegraphics[scale=0.47]{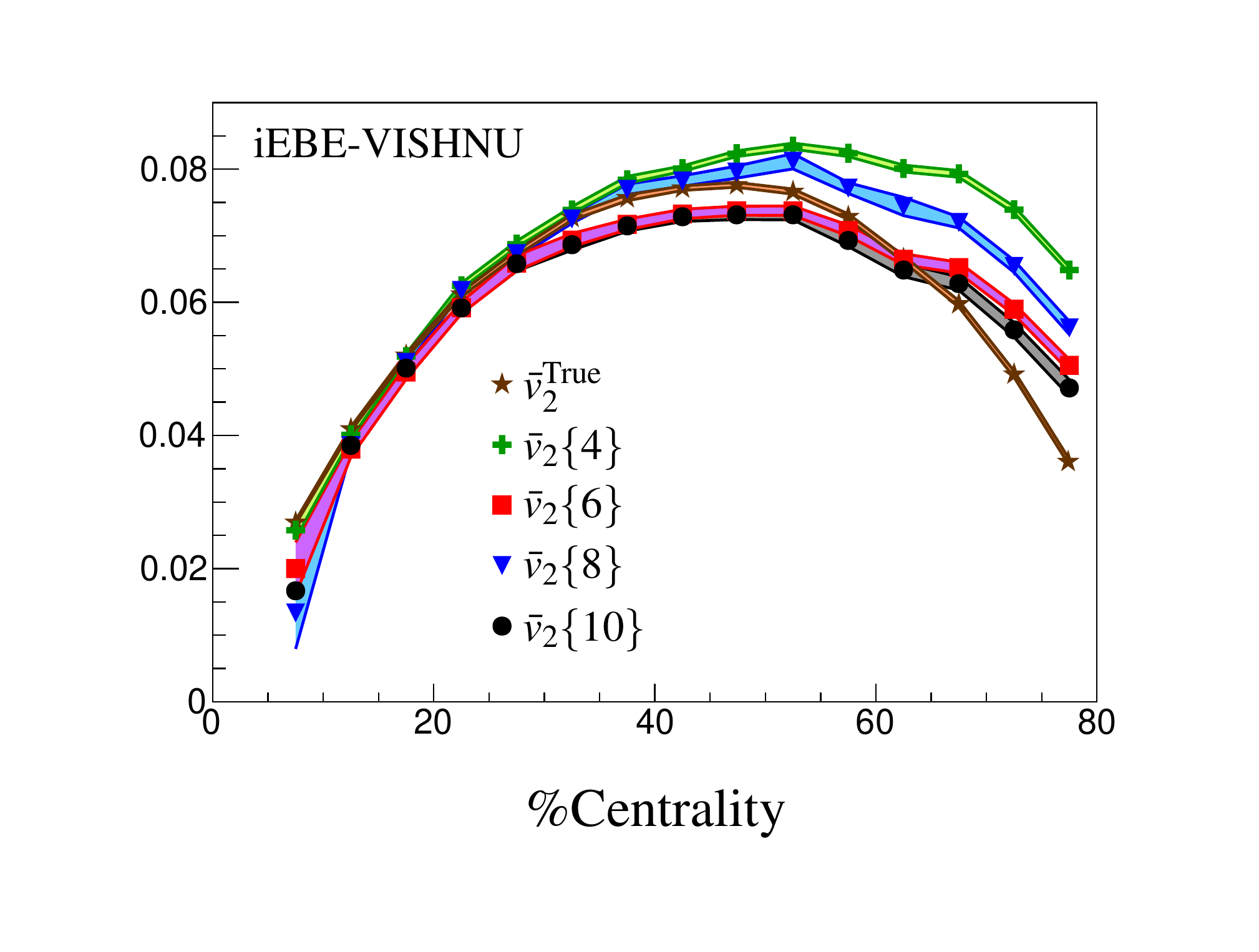} 			 	
		\caption{Using iEBE-VISHNU output, the true value of $\vb_2$ is compared with the estimators $\vb_2\{2k\}$.} 
		\label{checkEstims}
	\end{center}
\end{figure}

For $k=1$, we have $q_2\{2\}=0$ which means $\vb_2=v_2\{2\}$. For this special choice, all the $\Gamma_{2k-2}$ in  Eq.~\eqref{gammaDefi} diverge unless we set all other $q_2\{2k\}$ to zero too. As a result, this choice corresponds to the delta function for $p(v_{2,x},v_{2,y})$.

The first non-trivial choice is $q_2\{4\}=0$. Referring to Eq.~\eqref{newCumulsB}, we find
\bea\label{extimate4}
\vb_2\{4\}=v_2\{4\},
\eea
where in the above $\vb_2\{4\}$ refers to $\vb_2$ which is estimated from $q_2\{4\}=0$. This is exactly the assumption that has been made in Ref.~\cite{Giacalone:2016eyu} to find the skewness experimentally. By estimating $\vb_2$, we can find other $q_2\{2k\}$. We present a few first cumulants in the following,
\begin{subequations}\label{vbar4Cumuls}
	\begin{eqnarray}
	q_2\{4\}&=&0\\
	q_2\{6\}&=&-24\, v_2^5\{6\}\,\Delta_2\{4,6\}+\mathcal{O}(\Delta^{3/2})\\
	q_2\{8\}&=&-24\, v_2^7\{8\}\,\left(\Delta_2\{4,6\}-11\Delta_2\{6,8\}\right)\nn\\
	&&\hspace*{3.5cm}+\mathcal{O}(\Delta^{3/2})\\
	q_2\{10\}&=&240\, v_2^9\{10\}\,\left(3\Delta_2\{6,8\}-19\Delta_2\{8,10\}\right)\nn\\
	&&\hspace*{3.5cm}+\mathcal{O}(\Delta^{3/2}),
	\end{eqnarray}
\end{subequations}
where we used the notation 
\bea
\Delta_n\{2k,2\ell\}=v_n\{2k\}-v_n\{2\ell\}
\eea
for the fine-splitting between different $v_n\{2k\}$'s. Furthermore, in Eqs.~\eqref{vbar4Cumuls} we expanded $q_2\{2k\}$ in terms of fine-splitting $\Delta_2\{2k,2\ell\}$.

Note that the above $q_2\{2k\}$ are characterizing an estimated distribution $p(v_2;\vb_2\{4\})$. For such an estimated distribution, $q_2\{6\}$ is proportional to the skewness introduced in Ref.~\cite{Giacalone:2016eyu}. Interestingly, $q_2\{8\}$ is proportional to $\Delta_2\{4,6\}-11\Delta_2\{6,8\}$ which has been considered to be zero in Ref.~\cite{Giacalone:2016eyu}. However, here we see that this combination can be non-zero and its value is related to the cumulant $q_2\{8\}$. In fact, the same quantity can be computed  for a generic narrow distribution \cite{Jia:2014pza}. In turns out that this quantity can be non-vanishing in the small fluctuation limit.

The equations in \eqref{vbar4Cumuls} indicate that by assuming $q_2\{4\}=0$ all the other cumulants of $p(v_2;\vb_2\{4\})$ are written in terms of the fine-splitting $\Delta_2\{k,\ell\}$. Therefore, the distribution $p(v_2;\vb_2\{4\})$ satisfies all the fine-splitting structure  of $v_2\{2k\}$ by construction.

One can simply check that how much the estimation $\vb_2\{2k\}$ is accurate by using simulation. We exploit again the iEBE-VISHNU event generator to compare the true value of $\vb_2$ ($\vb_2^{\text{True}}$) with $\vb_2\{4\}=v_2\{4\}$.  The result is depicted in Fig.~\ref{checkEstims} by brown and green curves for $\vb_2^{\text{True}}$ and $\vb_2\{4\}$, respectively. As the figure illustrates, $\bar{v}_2\{4\}$ is not compatible with $\vb_2^{\text{True}}$ for centralities higher than $50\%$ where we expect that Bessel-Gaussian distribution does not work well. It should be noted that all other $v_2\{2k\}$ for $k>2$ never close to the true value of $\vb_2$ in higher centralities because all of them are 
very close to $v_2\{4\}$.

In order to improve the estimation of $\vb_2$, we set $q_2\{6\}=0$ in Eq.~\eqref{newCumulsC}. This equation has six roots where only two of them are real and positive.
In addition, as can be seen from Fig.~\ref{checkEstims}, the true value of $\vb_2$ is always smaller than $v_2\{4\}$ ($=\vb_2\{4\}$) in higher centralities. In fact, it is true for all $v_2\{2k\}$ for $k>2$. Based on this observation, we demand the root to be  smaller than $v_2\{4\}$. In fact, we have checked that the equation $q_2\{2k\}=0$ for $k=3,4,5$ has only one root which is real, positive and smaller than $v_2\{2k\}$ for $k=2,3,4,5$. 

In Fig.~\ref{checkEstims}, $\vb_2\{6\}$ is plotted by a red curve which is obtained by solving $q_2\{6\}=0$ numerically. As can be seen, it is more close to the real value of $\vb_2$ rather than $v_2\{4\}$.  In fact, we are able to find this root analytically too,
\bea\label{vb6Estim}
\vb_2\{6\}=v_2\{6\}-\sqrt{v_2\{6\}\Delta_2\{4,6\}}+\OO(\Delta),
\eea
which is compatible with the red curve in Fig.~\ref{checkEstims} with a good accuracy. Using the estimator \eqref{vb6Estim}, we find other $q_2\{2k\}$ as follows,
\begin{subequations}\label{q6Estm}
	\begin{eqnarray}
	q_2\{4\}&=&-4\, v_2^3\{6\}\,\sqrt{v_2\{6\}\Delta_2\{4,6\}}+\OO(\Delta),\\
	q_2\{6\}&=&0,\\
	q_2\{8\}&=&264\, v_2^7\{8\}\,\left(\Delta_2\{6,8\}-\Delta_2\{4,6\}\right)\nn\\
	&&\hspace*{3.5cm}+\mathcal{O}(\Delta^{3/2}),\\
	q_2\{10\}&=&240\, v_2^9\{10\}\,\left(3\Delta_2\{6,8\}-19 \Delta_2\{8,10\}\right)\nn\\
	&&\hspace*{3.5cm}+\mathcal{O}(\Delta^{3/2}).
	\end{eqnarray}
\end{subequations}
By comparing Eqs.~\eqref{vbar4Cumuls} and \eqref{q6Estm}, we note that $q_2\{2k\}$ (except $q_2\{10\}$) are different because the estimated distributions $p(v_2;\vb_2\{4\})$ and $p(v_2;\vb_2\{6\})$ are  different.

We can go further and estimate $\vb_2$ by solving the equation $q_2\{8\}=0$. The result is plotted by blue curve in Fig.~\ref{checkEstims}. Also, its analytical value can be found as follows,
\bea\label{vb8Estim}
\vb_2\{8\}&=& v_2\{8\}-\frac{1}{\sqrt{10}}\sqrt{v_2\{8\}\left(11 \Delta_2\{6,8\}-\Delta_2\{4,6\}\right)}\nn\\
&&\hspace*{3.5cm}+\mathcal{O}(\Delta).
\eea

As Fig.~\ref{checkEstims} indicates, comparing $\vb_2\{6\}$, the estimator $\vb_2\{8\}$  is a worse estimation of $\vb_2$ (except between the range $30\%$ to $50\%$ centralities). We might expect that because $q_2\{8\}\ll q_2\{6\}$  (see Fig.~\ref{qmqn}), the quantity $\vb_2\{8\}$ should be a better approximation than $\vb_2\{6\}$.  But this argument is not true. In fact, the cumulants $q_2\{2k\}$ for the true distribution are small, but they are non-zero at any centralities. Let us rewrite the Eqs.~\eqref{newCumulsB}-\eqref{newCumulsD} as follows,
\begin{subequations}\label{eqConstp}
	\begin{eqnarray}
	&&\vb_2^4-\delta_4v_2^4\{4\}=0,\label{eqConstApp}\\
	&&2\vb_2^6-6 v_2^4\{4\}\vb_2^2+4\,\delta_6\,v_2^6\{6\}=0,\label{eqConstAp}\\
	&&9\vb_2^8+6v_2^4\{4\}\vb_2^4-48 v_2^6\{6\} \vb_2^2+33\,\delta_8\,v_2^8\{8\}=0,\qquad \label{eqConstBp}
	\end{eqnarray}
\end{subequations}
where
\bea
\delta_{2k}=1-\frac{q_2\{2k\}}{c_2\{2k\}}.
\eea
The estimator $\vb_2\{2k\}$ ($k=2,3,4$) can be found by solving  the Eqs.~(\ref{eqConstp}a,b,c) where we set $\delta_{2k}=1$. Alternatively, by employing the actual value of $\delta_{2k}$ from the simulation we find $\vb_2^{\text{True}}$.
In fact, the difference between $\vb_2^{\text{True}}$  and $\vb_2\{2k\}$ is a manifestation of the inaccuracy in the setting $\delta_{2k}=1$. In other words, demanding $q_2\{2k\}=0$ is not exactly correct.
Looking at the problem from this angle, by referring to Fig.~\ref{checkEstims}, we realize that $\delta_4=1$ is the most inaccurate approximation. Also, $\delta_6=1$ is  more accurate than $\delta_8=1$. 

\begin{figure}[t!]
	\begin{center}
		\hspace*{-0.8cm}	\includegraphics[scale=0.47]{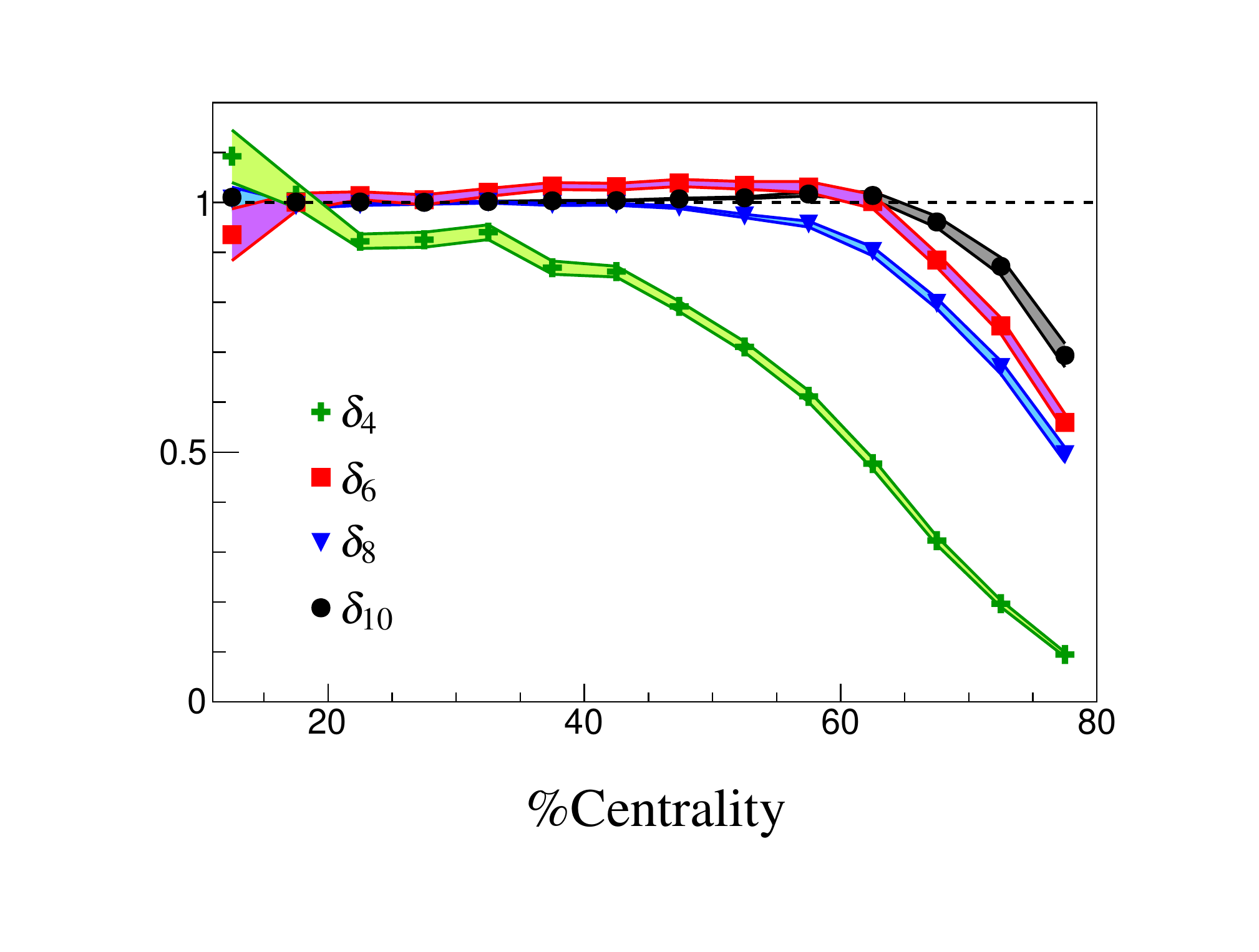} 	
		\caption{The quantity $\delta_{2k}$ for $k=2,3,4$ and $5$ with respect to centrality class.} 
		\label{estimApprox}
	\end{center}
\end{figure}

\begin{figure}[t!]
	\begin{center}
		\hspace*{-0.8cm}	 		\includegraphics[scale=0.47]{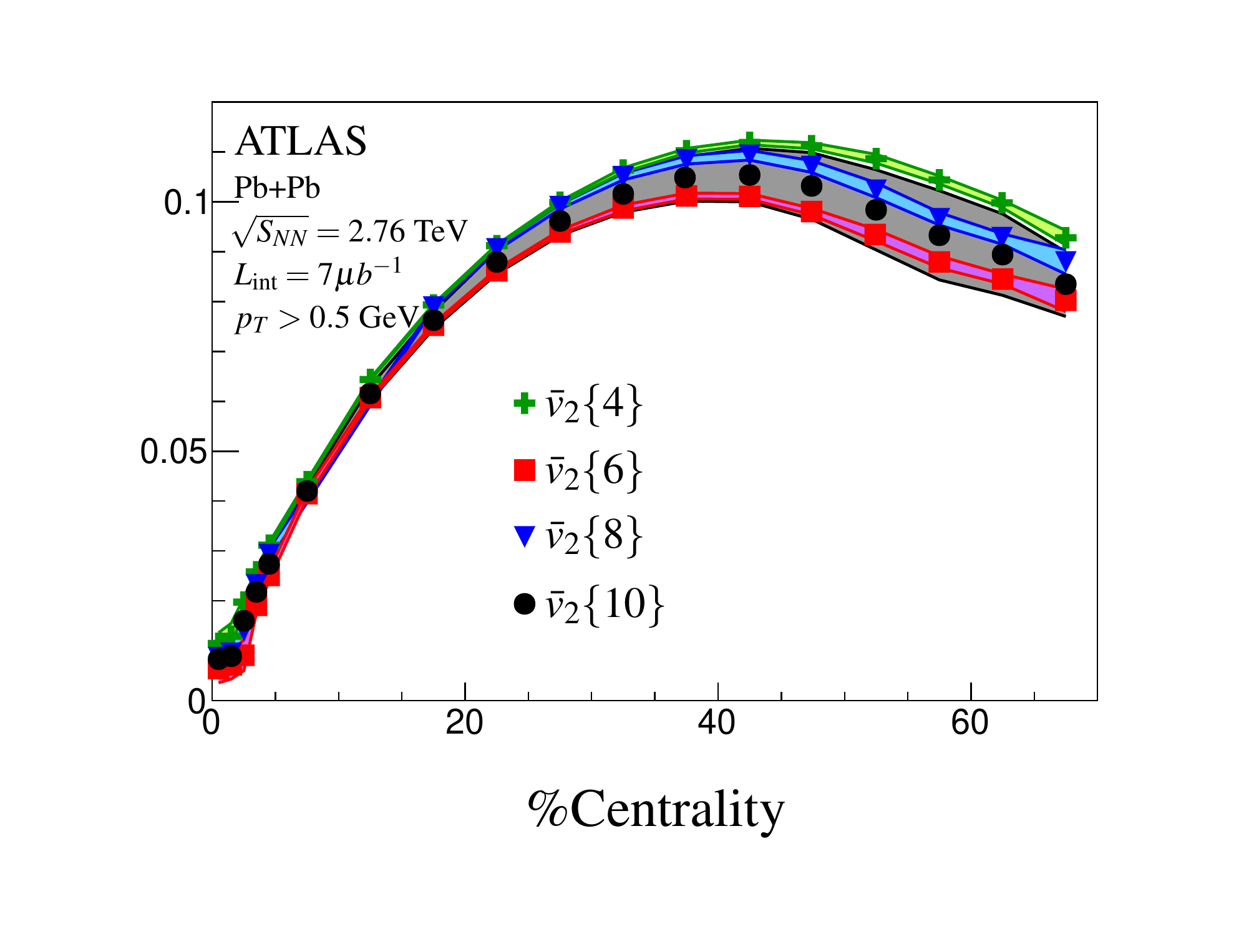} 		 	
		\caption{The averaged flow harmonic estimator $\vb_2\{2k\}$ obtained from the ATLAS experimental data \cite{Aad:2013xma}.} 
		\label{EstimsRealData}
	\end{center}
\end{figure}

By using iEBE-VISHNU generated data, we can check the accuracy of the $\delta_{2k}=1$ estimation by comparing different values of $\delta_{2k}$ (for $k=2,3,4,5$) calculated from simulation. The result is plotted in Fig.~\ref{estimApprox}. This figure confirms the difference between the estimators $\vb_2\{2k\}$ discussed above. The quantity $\delta_{4}$ has the most deviation from unity. Also, we see that $\delta_{8}$ deviates from unity for centralities above $55\%$ while $\delta_6$ (and $\delta_{10}$) is more close to one up to $65\%$ centrality. Moreover, $\delta_6$ ($\delta_{10}$) is larger than $\delta_8$ for all centralities. This can be considered as a reason for the fact that $\vb_2\{8\}$ is less accurate than $\vb_2\{6\}$ (and $\vb_2\{10\}$).

Furthermore, let us mention that the cumulant $q_2\{6\}$ changes its sign (see Fig.~\ref{qmqn} and Fig.~\ref{estimApprox}) for the centralities around $60$-$65\%$. It means it is exactly equal to zero at a specific point in this range, and we expect that $\vb_2\{6\}$ becomes exactly equal to $\vb_2^{\text{True}}$ at this point. This can be seen also in Fig.~\ref{checkEstims} where the red curve ($\vb_2\{6\}$) crosses the brown curve ($\vb_2^{\text{True}}$).

\begin{figure*}[t!]
	\begin{center}
		\begin{tabular}{ccc}
			\includegraphics[scale=0.4]{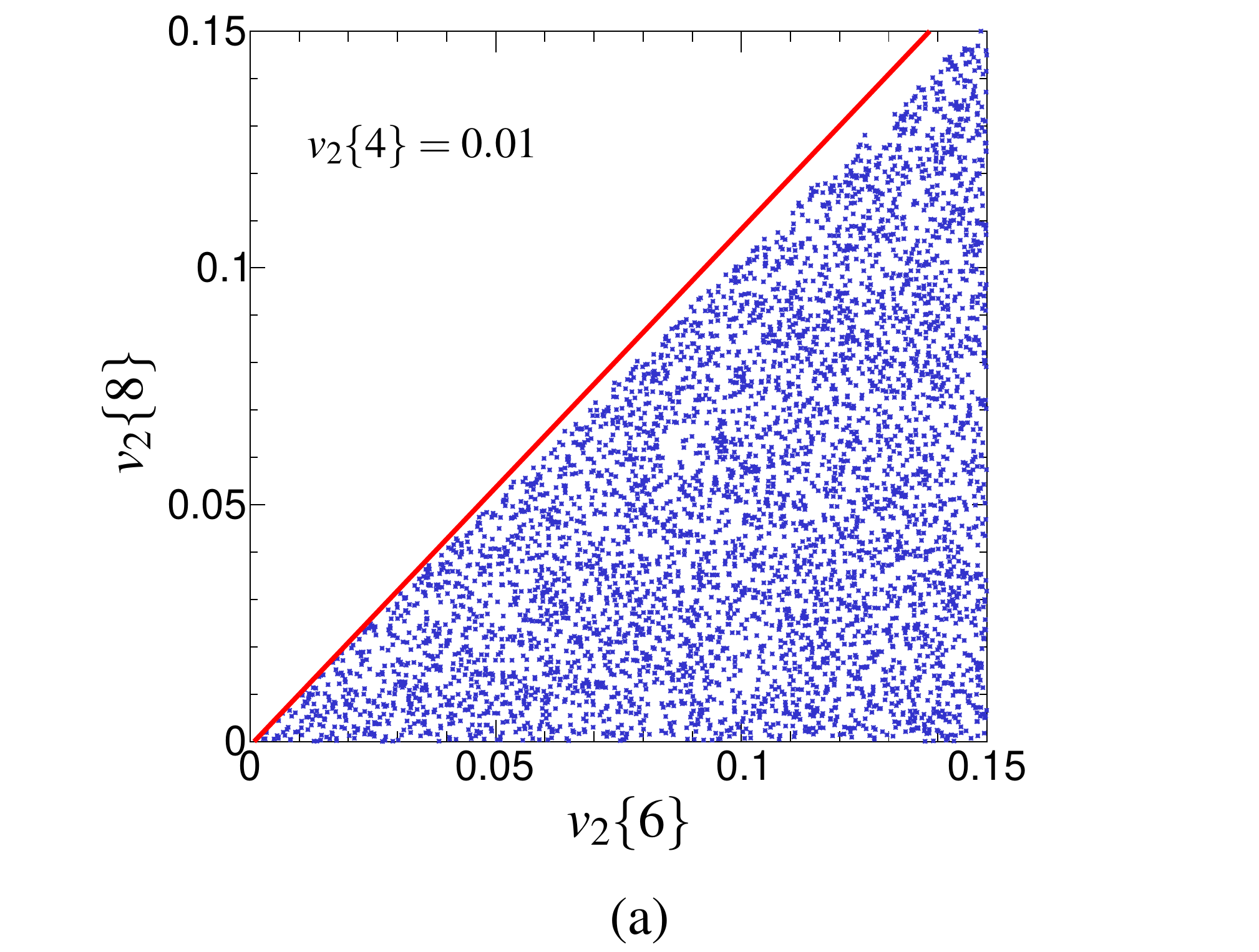} & &
			\includegraphics[scale=0.4]{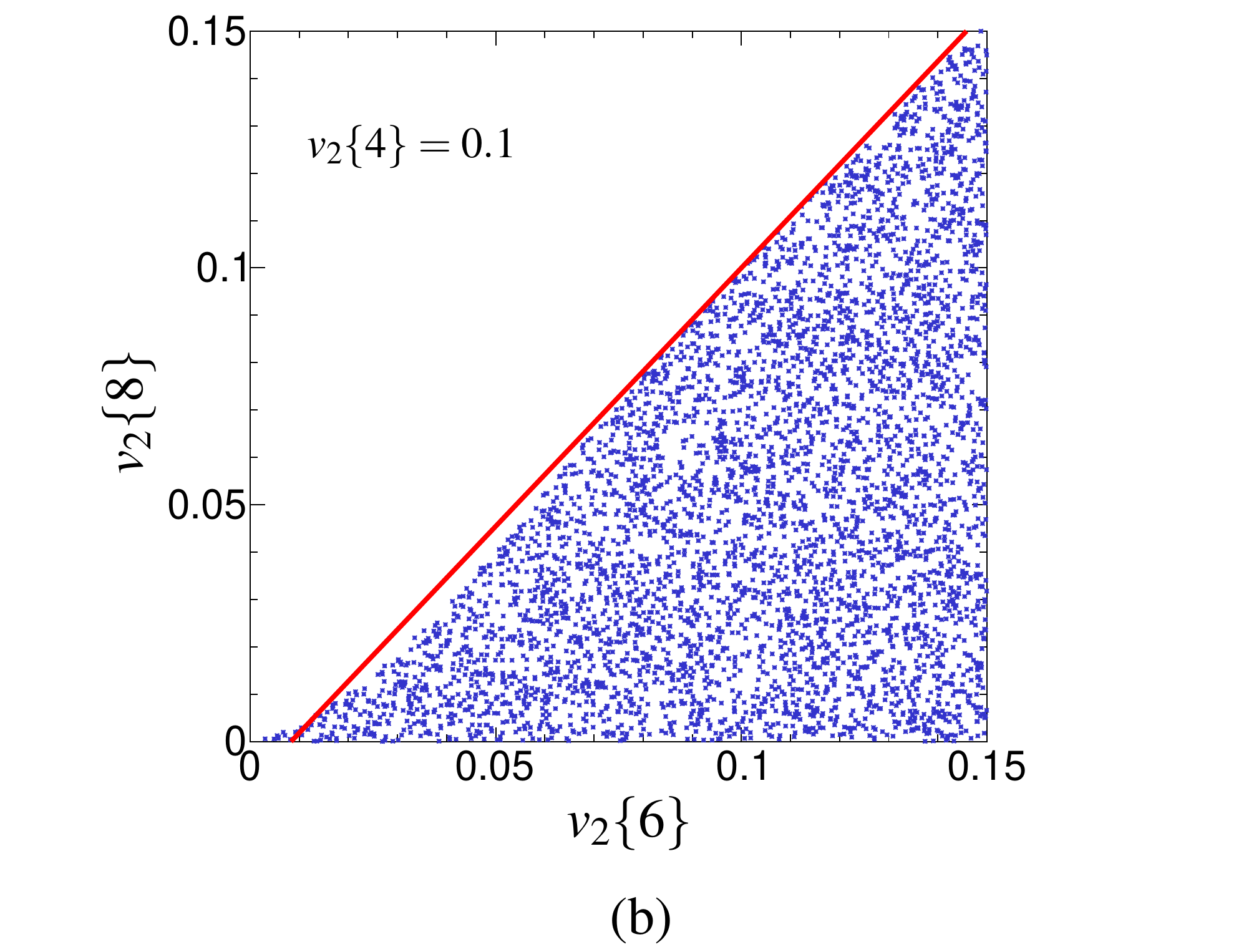}
		\end{tabular}		
		\caption{The allowed region of $v_2\{6\}$-$v_2\{8\}$ phase space for the two fixed values of $v_2\{4\}$ and $\delta_6=\delta_8=1$.} 
		\label{scatterPlot}
	\end{center}
\end{figure*}

The situation for $q_2\{10\}$ is very similar to $q_2\{6\}$. As a result, it is not surprising if we find the estimator $\vb_2\{10\}$ similar to $\vb_2\{6\}$. We have found $\vb_2\{10\}$ by solving $q_2\{10\}=0$ numerically. The result is plotted by a black curve in Fig.~\ref{checkEstims}. As can be seen, the results of $\vb_2\{6\}$ and $\vb_2\{10\}$ are approximately similar.

Now, we are in a position to estimate the $\vb_2$ of the real data by using $\vb_2\{2k\}$. According to the above discussions, we expect that $\vb_2\{6\}$, $\vb_2\{8\}$ and $\vb_2\{10\}$ are more close to the real value of the averaged ellipticity $\vb_2$ compared to $v_2\{4\}$ (or any other $v_2\{2k\}$ for $k>2$). The result is plotted in Fig.~\ref{EstimsRealData}. In finding the estimated $\vb_2 $, we employed  $v_2\{2k\}$ reported by ATLAS collaboration in Ref.~\cite{Aad:2013xma}. The value of $\vb_2\{4\}$ is exactly equal to $v_2\{4\}$ which is plotted by green curve in the figure. By plugging experimental values of $v_2\{2k\}$ into the relations \eqref{newCumulsC}, \eqref{newCumulsD} and \eqref{newCumulsE} and setting them to zero, we have numerically found $\vb_2\{6\}$ (red curve), $\vb_2\{8\}$ (blue curve) and $\vb_2\{10\}$ (black curve), respectively\footnote{In details, all the Eqs.~\eqref{newCumulsC}-\eqref{newCumulsE} were written in terms of moments $\la v_2^{2k}\ra$. Considering the reported experimental distribution $p(v_2)$ in Ref.~\cite{Aad:2013xma}, we are able to produce the covariance matrix associated with statistical fluctuations of the moments $\la v_2^{2k}\ra$. Using the covariance matrix, we generated 10000 random number by using a multidimensional Gaussian distribution. Employing each random number, we solved the equations \eqref{newCumulsC}-\eqref{newCumulsE} numerically and found estimated $\vb_2$. We obtained the standard deviation of the final $\vb_2$ distribution as the statistical error of the $\vb_2\{2k\}$.}. The errors of $\vb_2\{10\}$ is too large for the present experimental data, and more precise observation is needed to find more accurate estimation.  Exactly similar to the iEBE-VISHNU simulation, the value of $\vb_2\{8\}$ is between $v_2\{4\}$ and $\vb_2\{6\}$\footnote{We have computed the quantities $\bar{\varepsilon}_2\{4\}$, $\bar{\varepsilon}_2\{6\}$ and $\bar{\varepsilon}_2\{8\}$ for Elliptic-Power distribution \cite{Yan:2014afa} which is a simple analytical model for the initial state distribution. We have observed exactly the same hierarchy, and we have seen $\bar{\varepsilon}^{\text{True}}$ is more close to $\bar{\varepsilon}_2\{6\}$. This is an evidence that this behavior is generic for the interested distributions in heavy ion physics. }. Therefore, we expect the true value of the averaged ellipticity to be close to the value of $\vb_2\{6\}$\footnote{Comparing Fig.~\ref{checkEstims} with Fig.~\ref{EstimsRealData}, one finds that the values of $\vb_2\{2k\}$ from simulation are relatively smaller than that obtained from the real data. This deviation is due to the difference in $p_T$ range. In Fig.~\ref{EstimsRealData}, we used the data from  Ref.~\cite{Aad:2013xma} where $p_T>0.5\,\text{GeV}$, while the output of the iEBE-VISHNU is in the range $p_T \lesssim 4\,\text{GeV}$. For a confirmation of iEBE-VISHNU output, we refer the reader to Ref.~\cite{Sirunyan:2017fts,Acharya:2018lmh} where $v_2\{4\}$ is reported for $p_T$ below $3\,\text{GeV}$. The order of magnitude of $\vb_2\{2k\}$ in our simulation is compatible with that mentioned in Ref.~\cite{Sirunyan:2017fts,Acharya:2018lmh}.}.

In this section, we introduced a method to estimate $p(v_2,\vb_2)$. By considering the cumulants $c_2\{2k\}$ of the true distribution $p(v_2)$, we estimated $\vb_2$ by assuming that the cumulant $q_2\{2k\}$ of $p(v_2,\vb_2)$ is zero for a specific value of $k$. These estimations for $q_2\{6\}=0$ and $q_2\{8\}=0$ are presented analytically in Eq.~\eqref{vb6Estim} and Eq.~\eqref{vb8Estim} and also with red and blue curves in Fig.~\ref{checkEstims} numerically. We exploited hydrodynamic simulation to investigate the accuracy of our estimations. We found that $\vb_2\{6\}$ is more accurate than $\vb_2\{8\}$. Finally, we found the experimental values for $\vb_2\{6\}$, $\vb_2\{8\}$ and $\vb_2\{10\}$. 

Until now, we considered the cumulants $v_n\{2k\}$ as an input to find an estimation for $\vb_n$. In the next section, we try to restrict the phase space of the allowed region of $v_n\{2k\}$ by using cumulants $q_n\{2k\}$.

\section{Constraints On the Flow Harmonics Phase Space}\label{secV}

Referring to Eq.~\eqref{vb6Estim} and Eq.~\eqref{vb8Estim}, we see that these estimators lead to real values for $\vb_2\{4\}$ and $\vb_2\{6\}$ only if we have $11 \Delta_2\{6,8\}\geq \Delta_2\{4,6\}\geq 0$. In this section, we would like to investigate these constraints and their validity range.

We first consider the Eq.~\eqref{eqConstAp}. The quantities $v_2\{2k\}$ and $\vb_2$ are real valued. Therefore, it is a well-defined and simple question that what are the allowed values of $v_2\{4\}$ and $v_2\{6\}$ such that the  Eq.~\eqref{eqConstAp} has at least one real root. The polynomial in the left-hand side of Eq.~\eqref{eqConstAp} goes to positive infinity for $\vb_2\to \pm \infty$. As a result, it has at least one real root if the polynomial is negative in at least one of its minima. This condition is satisfied for
\bea\label{exactIneqI}
\delta_6^{1/6} v_2\{6\} \leq v_2\{4\}.
\eea 
Since $\delta_6$ is unknown, there is no bound on $v_2\{4\}$ and $v_2\{6\}$. Although we do not know the exact value of $\delta_6$, we know that $0.9 \lesssim \delta_6^{1/6} \lesssim 1$ based on our simulation (see Fig.~\ref{estimApprox}). In this case, if we observe $v_2\{6\} \leq v_2\{4\}$, we immediately deduce the inequality in Eq.~\eqref{exactIneqI}. On the other hand, there is a possibility to observe that $v_2\{6\}$ is slightly greater than $v_2\{4\}$. This observation means that $\delta_6$ is definitely smaller than unity. Here, we show both cases by an approximate inequality as $v_2\{4\}\gtrsim v_2\{6\}$ due to the smallness of $\delta_6^{1/6}$.

\begin{figure}[t!]
	\begin{center}
		\begin{tabular}{c}
			\includegraphics[scale=0.4]{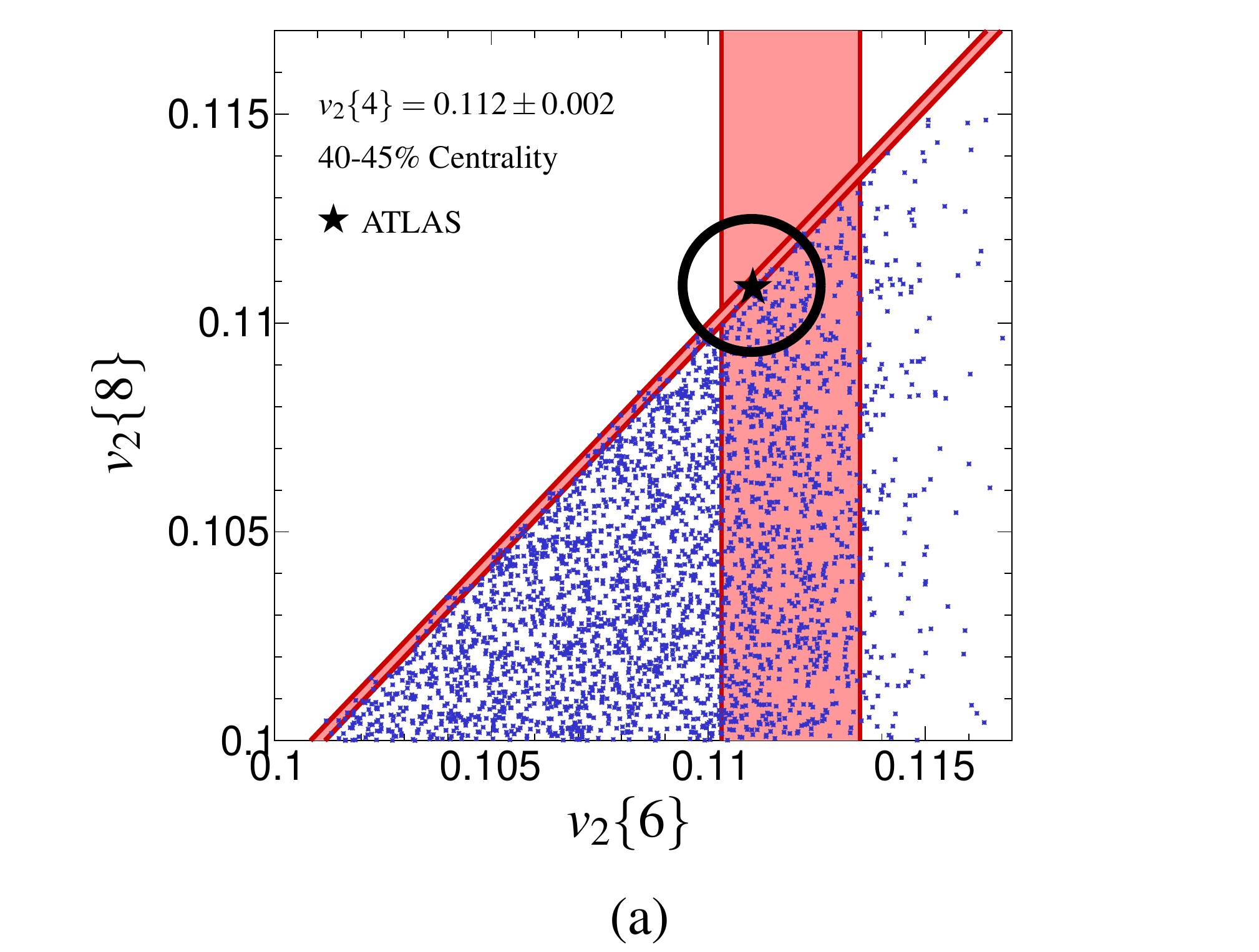}\\
			\\
			\includegraphics[scale=0.4]{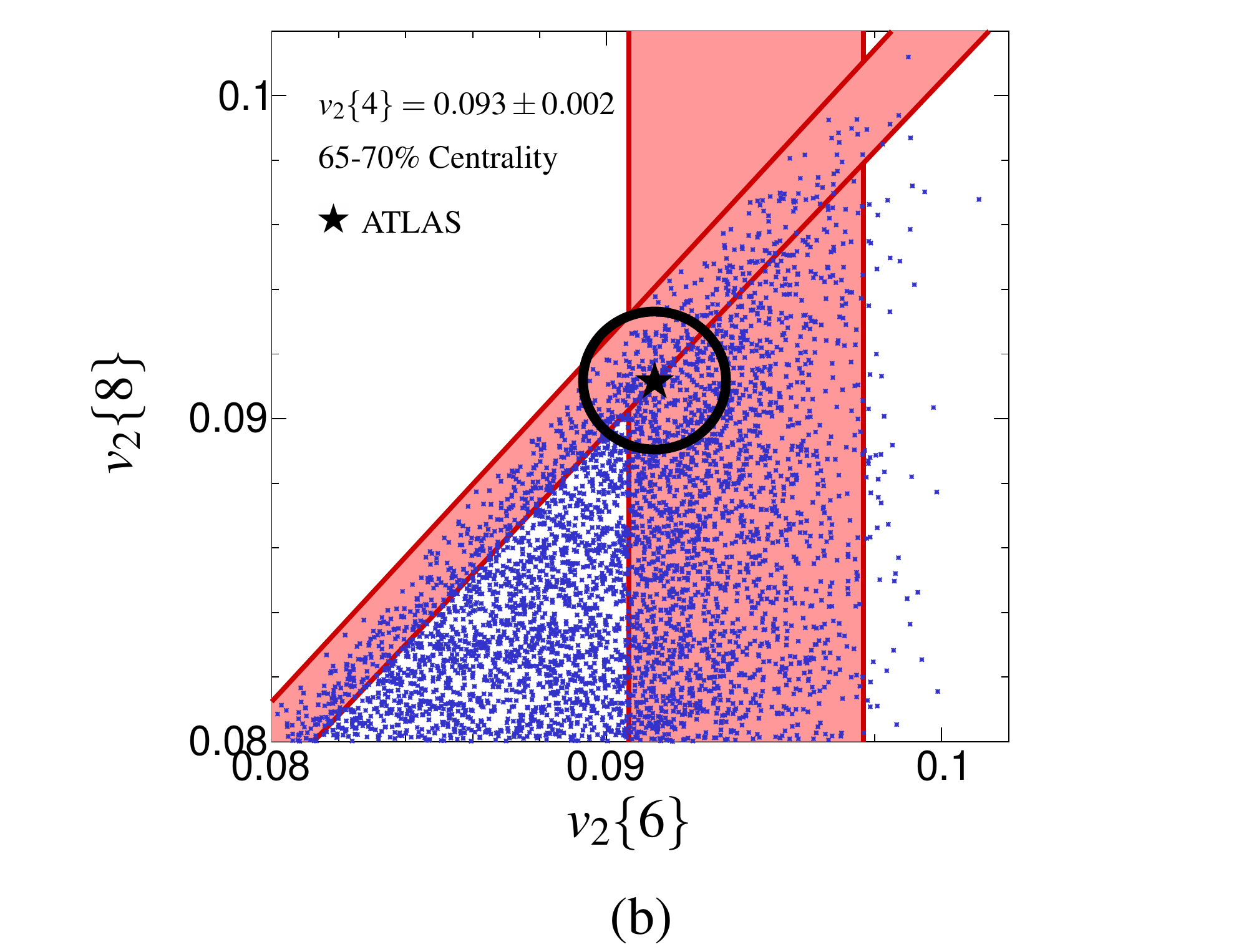} 			 			
		\end{tabular}		
		\caption{Combining the constraints $\Delta_2\{4,6\} \gtrsim 0$ and $11 \Delta_2\{6,8\}-\Delta_2\{4,6\} \gtrsim 0$ to find the allowed region of $v_2\{6\}$-$v_2\{8\}$ phase space. In this case, $\delta_6$, $\delta_6$ and $v_2\{4\}$ are not completely fixed. The region is compatible with the ATLAS data reported in Ref.~\cite{Aad:2013xma}. } 
		\label{DomainCheck}
	\end{center}
\end{figure}

Alternatively, it is known that the initial eccentricity point $(\ve_{2,x},\ve_{2,y})$ is bounded into a unit circle \cite{Yan:2014afa}, and it leads to a negative skewness for $p(\ve_{2,x},\ve_{2,y})$ in non-central collisions. By considering the hydrodynamic linear response, the skewness in $p(\ve_{2,x},\ve_{2,y})$  is translated into $p(v_{2,x},v_{2,y})$ skewness and condition $v_2\{4\}>v_2\{6\}$ \cite{Giacalone:2016eyu}. However, it is possible that the non-linearity of the hydrodynamic response changes the order in the inequality to the case that $v_2\{6\}$ is slightly greater than $v_2\{4\}$. This is compatible with the result which we have found from a more general consideration.

Now, we concentrate on the Eq.~\eqref{eqConstBp}. Due to the complications in finding the analytical allowed values of $v_2\{2k\}$, we investigate it numerically. First, we consider the case that $\delta_6=\delta_8=1$. In this case, we fix a value for $v_2\{4\}$ and after that randomly generate $v_2\{6\}$ and $v_2\{8\}$ between $0$ to $0.15$. Putting the above generated and fixed values into Eq.~\eqref{eqConstBp}, we can find $\vb_2$ numerically. If the equation has at least one real solution, we accept $(v_2\{6\},v_2\{8\})$, otherwise we reject it. The result is presented as scatter plots in Fig.~\ref{scatterPlot}. As can be seen from the figure, some region of the $v_2\{2k\}$ phase space is not allowed. The condition $11 \Delta_2\{6,8\}\geq \Delta_2\{4,6\}$ (see the square root in \eqref{vb8Estim}) indicates that the border of this allowed region can be identified with $v_2\{8\}=(12v_2\{6\}-v_2\{4\})/11$ up to order $\Delta_2\{2k,2\ell\}$. This is shown by red line in Fig.~\ref{scatterPlot}. Alternatively, the numerically generated border of the allowed region slightly deviates from the analytical border line. It happens for the region that $v_2\{4\}$ is different from $v_2\{6\}$ and $v_2\{8\}$ considerably. The reason is that $\Delta_2\{2k,2\ell\}$ is not small in this region, and the condition $11 \Delta_2\{6,8\}\geq \Delta_2\{4,6\}$ is not accurate anymore.

Let us combine the constraint obtained from Eq.~\eqref{eqConstAp} and Eq.~\eqref{eqConstBp}. For a more realistic study, we use the ATLAS data for $v_2\{4\}$ as an input. Instead of using a fixed value for $v_2\{4\}$, we generate it randomly with a Gaussian distribution where it is centered  around the central value of $v_2\{4\}$, and the width equal to the error of $v_2\{4\}$. The result for $40$-$45\%$ centralities is presented in Fig.~\ref{DomainCheck}(a). For this case, we expect that the Bessel-Gaussian distribution works well. As a result, we assume $\delta_6\simeq \delta_8\simeq 1$ (see Fig.~\ref{estimApprox}). From ATLAS results \cite{Aad:2013xma}, we have $v_2\{4\}=0.112\pm 0.002$ in $40$-$45\%$ centralities. The black star in the figure shows the experimental value of $(v_2\{6\},v_2\{8\})$ (the ellipse shows the one sigma error without considering the correlations between $v_2\{6\}$ and $v_2\{8\}$).  
The width of the bands is due to the one sigma error of $v_2\{4\}$. As the figure shows, the experimental result is compatible with the allowed region of $(v_2\{6\},v_2\{8\})$.

For more peripheral collisions, we expect a non-zero value for $\delta_{2k}$. In the $65$-$70\%$ centrality class (the most peripheral class of events reported by ATLAS in Ref.~\cite{Aad:2013xma}), we have $v_2\{4\}\simeq 0.093\pm 0.002$. According to our simulation in this centrality class, we expect  that the values of  $\delta_6 $ and  $\delta_8$ to be $0.88$ and $0.8$, respectively. However, here we do not choose fixed values for $\delta_6$ and $\delta_8$. Instead, we generate a random number between $0.8$ to $1$ and assign the result to both $\delta_6$ and $\delta_8$. The result is presented in Fig.~\ref{DomainCheck}(b). Referring to this figure, the allowed region is compatible with the experiment similar to the previous case. 
For non-zero values of $\delta_6$ and $\delta_8$, the allowed region can be identified by $v_2\{6\}=\delta_6^{-1/6} v_2\{4\}$ and $v_2\{8\}=\delta_8^{-1/8}(12v_2\{6\}-v_2\{4\})/11$. These two constraints (similar to Fig.~\ref{DomainCheck}(a)) are presented by two bands in Fig.~\ref{DomainCheck}(b). In this case, the width of the bands is due to the inaccuracy in $\delta_6$ and $\delta_8$ together with $v_2\{4\}$.

By considering the correlation between $v_2\{6\}$ and $v_2\{8\}$, the experimental one sigma region of $v_2\{6\}$-$v_2\{8\}$ would not be a simple domain. Nevertheless, for the present inaccurate case which is depicted in Fig.~\ref{DomainCheck}, 
we are able to restrict the one sigma domain by comparing it with the allowed region showed by the blue dots.

Let us summarize the constraint on the flow harmonic fine-splitting as follows
\begin{subequations}\label{finSplitCons}
	\begin{eqnarray}
	&&\Delta_2\{4,6\} \gtrsim 0, \label{finSplitConsA}\\
	&&11 \Delta_2\{6,8\}-\Delta_2\{4,6\} \gtrsim 0,\label{finSplitConsB}
	\end{eqnarray}
\end{subequations}
which correspond to the region filled by the blue dots in Fig.~\ref{DomainCheck}. Note that we used the approximate inequality in Eq.~\eqref{finSplitConsB} because of not only ignoring $\delta_8^{1/8}$ but also ignoring terms with order $\OO(\Delta^2)$.

\begin{figure}[t!]
	\begin{center}
		\hspace*{-0.5cm}	\includegraphics[scale=0.45]{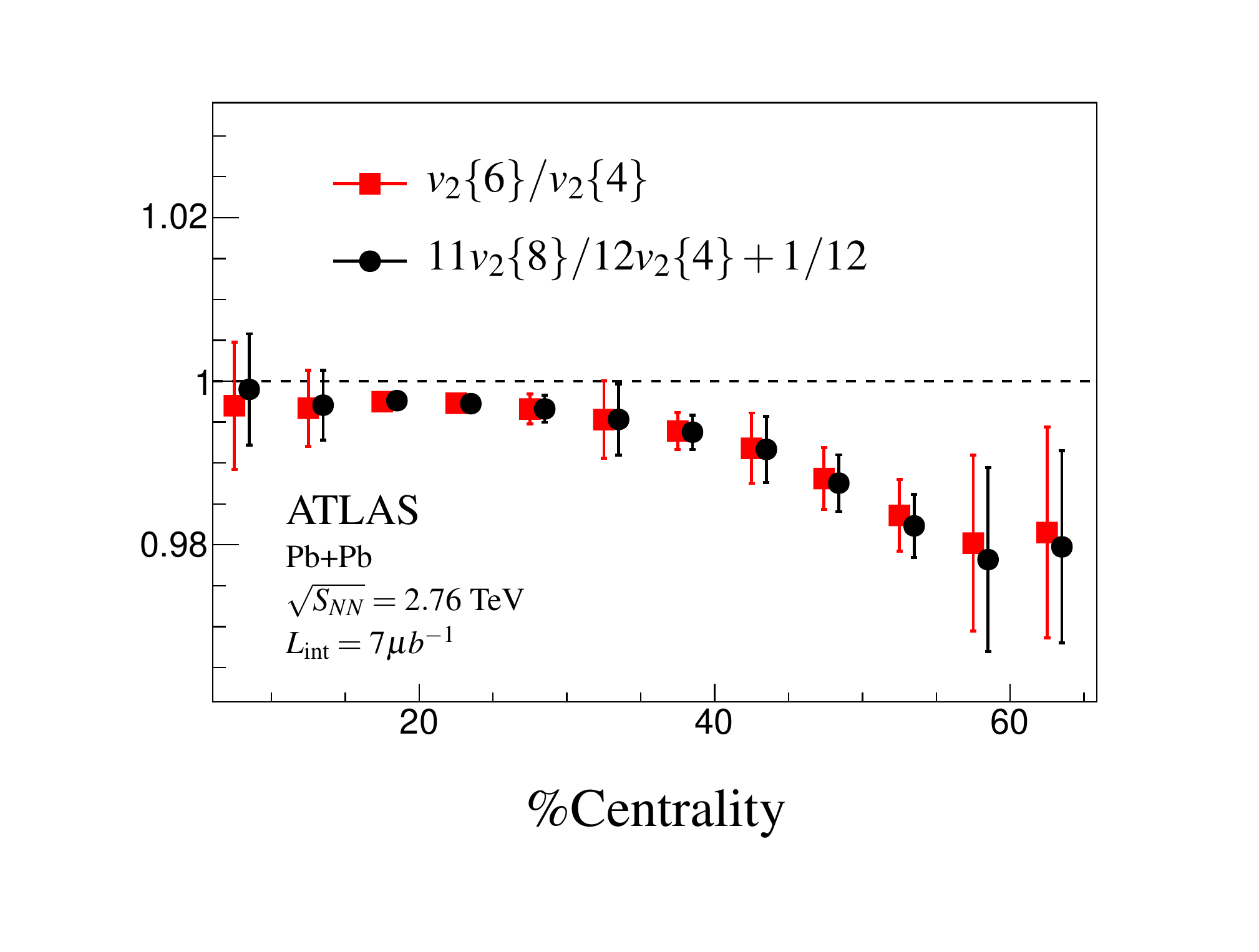} 			 	
		\caption{Experimental values of ratios $v_2\{6\}/v_2\{4\}$ and $11v_2\{6\}/12v_2\{4\}+1/12$ with respect to the centrality class from ATLAS data \cite{Aad:2013xma}. } 
		\label{DataSpliting}
	\end{center}
\end{figure}

Let us point out that the inequalities in Eq.~\eqref{finSplitCons} can be written as
\bea\label{finSplitConsII}
\frac{11}{12} \frac{v_2\{8\}}{v_2\{4\}}+\frac{1}{12}\lesssim \frac{v_2\{6\}}{v_2\{4\}} \lesssim 1.
\eea
In Ref.~\cite{Aad:2013xma}, the ratios $v_2\{6\}/v_2\{4\}$ and $v_2\{8\}/v_2\{4\}$ have been calculated experimentally. In Fig.~\ref{DataSpliting}, we compared the relations $v_2\{6\}/v_2\{4\}$ with $11v_2\{6\}/12v_2\{4\}+1/12$. As can be seen, the conditions in Eq.~\eqref{finSplitConsII} are satisfied in all centralities. Furthermore, the inequality in the right-hand side of Eq.~\eqref{finSplitConsII} is not saturated while we need a more accurate experimental observation to decide about the saturation of the inequality in the left-hand side.

Finally we would like to mention that the Eq.~\eqref{newCumulsE} does not lead to any constraint on  $v_2\{2k\}$ for $0.5 \lesssim \delta_{10} \lesssim 1$. The reason is that $\vb_2=0$ is a minimum of the following relation
\begin{subequations}\nn
	\begin{eqnarray}
	&&156\, \vb_2^{10}-480\,v_2^4\{4\} \vb_2^6-120\, v_2^6\{6\} \vb_2^4\nn\\
	&&\hspace*{1cm}+20(12 v_2^8\{4\}+33v_2^8\{8\})\vb_2^2-456\,\delta_{10} v_2^{10}\{10\},\nonumber
	\end{eqnarray}
\end{subequations}
and its value in the minimum is $-456 \,\delta_{10} v_2^{10}\{10\}$. It means that for $\delta_{10}>0$ the equation always has real roots.

\section{Conclusion and Outlook}\label{secVI}

In the present work, we have employed the concept of Gram-Charlier A series to relate the distribution $p(v_n)$ to $c_n\{2k\}$. We have found an expansion around the Bessel-Gaussian distribution where the coefficients of the expansion have been written in terms of a new set of cumulants $q_n\{2k\}$. We have shown that the corrected Bessel-Gaussian distribution can fit the actual distribution $p(v_n)$ much better than Bessel-Gaussian distribution. The new cumulants $q_n\{2k\}$ was written in terms of $c_n\{2k\}$ and averaged flow harmonic $\vb_n$. Because the only non-vanishing new cumulants are $q_n\{2\}$ for Bessel-Gaussian distribution, they are more natural choice to study the distributions near Bessel-Gaussian compared to $c_n\{2k\}$.

By using the cumulants $q_n\{2k\}$, we could systematically introduce different estimations for $p(v_n)$ and consequently relate the averaged ellipticity $\vb_2$ to the flow harmonic  fine-splitting $v_2\{2k\}-v_2\{2\ell\}$ for $k,\ell \geq2$ and $k\neq \ell$. As an specific example for the $\vb_2$ estimator, we have shown that $\vb_2\{6\}\simeq v_2\{6\}-\sqrt{v_2\{6\}(v_2\{4\}-v_2\{6\})}$. We have used iEBE-VISHNU event generator to compare the true value of the $\vb_2$ to the estimated one, also, we have shown that the estimator $\vb_2\{6\}$ is more accurate than $v_2\{2k\}$ for $k>1$. As another application of new cumulants, we have constrained the phase space of the flow harmonics $v_n\{2k\}$ to the region that $v_2\{4\}-v_2\{6\}\gtrsim 0$ and $12 v_2\{6\}-11v_2\{8\}-v_2\{4\}\gtrsim 0$. It is experimentally confirmed that $v_2\{4\}-v_2\{6\}>0$. But we need more accurate experimental observation for the quantity $12 v_2\{6\}-11v_2\{8\}-v_2\{4\}$.

One should note that we have shown the compatibility of allowed phase space of $v_2\{2k\}$ with experimental results of (high multiplicity) Pb-Pb collisions. Recently, the flow harmonics are measured for p-p, p-Pb and low-multiplicity Pb-Pb collisions by ATLAS in Ref.~\cite{Aaboud:2017acw}. In the light of $q_2\{2k\}$ cumulants, it would be interesting to study the similarity and difference between the splitting of $v_2\{2k\}$ in these systems  and examine the compatibility of the results with the allowed region comes from $q_2\{2k\}$.

Furthermore, we have only focused on the distribution $p(v_n)$ in the present study. However, based on the  observation of symmetric cumulants and event-plane correlations, we expect that a similar systematic study for distribution $p(v_1,v_2,\ldots)$ can connect this joint distribution to the observations. Such a study would be helpful to relate the initial state event-by-event fluctuations to the observation. This would be a fruitful area for further work.

\section*{Acknowledgements}
The authors would like to specially thank Hessamaddin Arfaei, HM's supervisor, for useful discussions, comments and providing guidance over HM's work during this project. We would like to thank Jean-Yves Ollitrault for useful discussions and comments during ``IPM Workshop on Collective Phenomena \& Heavy Ion Physics''.  We also thank Ali Davody for providing us the iEBE-VISHNU data and discussions. We thanks Mojtaba Mohammadi Najafabadi and Ante Bilandzic for useful comments. We would like to thank Navid Abbasi, Davood Allahbakhshi, Giuliano Giacalone, Reza Goldouzian and Farid Taghinavaz for discussions. We thank participants of ``IPM Workshop on Collective Phenomena \& Heavy Ion Physics''.

\appendix

\section{Two Dimensional Cumulants in Cartesian and Polar Coordinates}\label{AppA}

The Cartesian cumulants $\A_{mn}$ can be found in terms of the moments $\la x^{k} y^{\ell} \ra$ by using the first line of the Eq.~\eqref{2DCumuls}. A few first cumulants can be found as follows,
\begin{subequations}
	\begin{eqnarray}
	\A_{10}&=& \la x\ra, \\
	\A_{01}&=& \la y\ra,\\
	\A_{20}&=& \la x^2\ra-\la x\ra^2,\\
	\A_{11}&=& \la x y\ra-\la x\ra \la y\ra,\\
	\A_{02}&=& \la y^2\ra-\la y\ra^2,\\
	\A_{30}&=& \la x^3\ra-3\la x\ra \la x^2\ra+2\la x\ra^3,  \\
	\A_{21}&=& \la x^2 y\ra-\la x^2\ra\la y \ra-2 \la x\ra \la x y\ra+2\la x\ra^2 \la y\ra,   \nn\\
	\\
	\A_{12}&=&  \la x y^2\ra-\la x \ra\la y^2\ra-2  \la x y\ra\la y\ra+2\la x\ra\la y\ra^2,   \nn\\
	\\
	\A_{03}&=& \la y^3\ra-3\la y\ra \la y^2\ra+2\la y\ra^3.
	\end{eqnarray}
\end{subequations}
Note that for a fixed number $\chi$ we have $\A_{mn} \to \chi^{m+n} \A_{mn}$ by replacing $x\to \chi x $ and $y\to \chi y $. We call $m+n$ as the order of $\A_{mn}$ . Consequently, there are $m+n+1$ number of cumulants with the order $m+n$.

In order to find the cumulants $\C_{m,n}$ in polar coordinates $x=r\cos\varphi$ and $y=r\sin\varphi$, we use the second line of the Eq.~\eqref{2DCumuls}. Considering the Jacobi-Anger identity
\bea
e^{i \mathbf{r}\cdot \mathbf{k}}=e^{i r k \cos(\varphi-\varphi_k)}=\sum_{n=-\infty}^{\infty}i^n J_n(kr)\,e^{in(\varphi-\varphi_k)},\qquad 
\eea
we are able to write the Eq.~\eqref{2DCumuls} as follows,
\begin{equation}\label{polarCumul}
\begin{aligned}
&\log \left(\sum_{m=0}^{\infty}\sum_{n=-\infty}^{\infty} \frac{(ik)^{2m+n} e^{-in\varphi_k}}{2^{2m+n}m!(m+n)!}\la r^{2m+n} e^{in\varphi}\ra\right)\\
&\hspace*{3cm}=\sum_{m=0}^{\infty}\sum_{n=-\infty}^{\infty} \frac{\C_{m,n}\; (ik)^m e^{in\varphi_k}}{m!},
\end{aligned}
\end{equation}
where in the left-hand side of the above we have used the series form of the Bessel function $J_n(kr)$.

In Eq.~\eqref{polarCumul}, the combination $(ik)^{2m+n}e^{-in\varphi_k}$ is appeared in the left-hand side. It means that if we have odd $|n|$ then the power of $ik$ is odd and if we have even $|n|$ then the power of $ik$ is even. Therefore, the only non-vanishing $\C_{m,n}$ are those both $m$ and $|n|$ are odd or even. The other consequence of the combination $(ik)^{2m+n}e^{-in\varphi_k}$ is that we have $\C_{m,n}=0$ for $|n|>m$. The reason is that in the right-hand side of the Eq.~\eqref{polarCumul}, the combination $(ik)^{m'} e^{i n' \varphi_k}$ is appeared. Therefore, in order to have a non-vanishing $\C_{m,n}$ we need to have some terms in the left-hand side such that $2m+n=m'$ and $n=-n'$. It immediately leads to $m=(m'+n')/2\geq0$ and $m+n=(m'-n')/2\geq0$. One should note that the $(m+n)!$ in the denominator of Eq.~\eqref{polarCumul} diverges if $m+n<0$. As a result ,we deduce that $\C_{m',n'}$ can be non-vanishing if $|n'|\leq m'$. Strictly speaking, for $m\geq 0$ only cumulants
\bea\label{CmnConst}
\C_{m,-m},\quad \C_{m,-m+2},\quad \cdots \quad \C_{m,m-2},\quad \C_{m,m}
\eea
are non-zero. Also, note that we have $\C_{m,-n}=\C_{m,n}^*$ because the distribution is a real function.

Based on the above considerations one can find all non-trivial values of $\C_{m,n}$ in terms of moments $\la r^{m} e^{in\varphi}\ra$ by using Eq.~\eqref{polarCumul}. A few first two dimensional cumulants in polar coordinates are presented explicitly in the following
\begin{subequations}\label{polarCumulantExampl}
	\begin{eqnarray}
	\C_{1,1}&=&\frac{1}{2}\la r e^{-i \varphi}\ra, \\
	\C_{2,0}&=&\frac{1}{2}\left[\la r^2\ra-\la r e^{i\varphi}\ra\la r e^{-i\varphi}\ra\right], \\
	\C_{2, 2}&=&\frac{1}{4}\left[\la r^2\,e^{ -2i \varphi}\ra-\la r\,e^{ -i \varphi}\ra^2\right], \\
	\C_{3, 1}&=&\frac{3}{8}\left[\la r^3 e^{-i\varphi}\ra+2\la r e^{i\varphi}\ra\la r e^{ -i\varphi}\ra^2\right.\nn\\
	&&\left.\hspace*{0.8cm}-\la r e^{i\varphi}\ra\la r^2 e^{ -2i\varphi}\ra-2\la r^2\ra\la r e^{ -i\varphi}\ra\right],\\
	\C_{3,3}&=&\nn\\
	&&\hspace*{-0.7cm}\frac{1}{8}\left[\la r^3 \,e^{ -3i \varphi}\ra-3\la r \,e^{ -i \varphi}\ra\la r^2 \,e^{ -2i \varphi}\ra+2\la r \,e^{ -i \varphi}\ra^3\right].\quad\;\;
	\end{eqnarray}
\end{subequations}
The first cumulant, $\C_{0,0}$, is equal to zero by considering the normalization condition of the probability distribution. 

By redefinition 
\[W_{0,n}=\C_{n,0},\quad W^c_{n,m}=\Re(\C_{m,n}), \quad  W^s_{n,m}=-\Im(\C_{m,n}),\]
we find the cumulants in polar coordinate which has been obtained in Ref.~\cite{Teaney:2010vd}. In Ref.~\cite{Teaney:2010vd}, the translational and rotational invariance have been used to eliminate $W^c_{1,1}$, $W^s_{1,1}$ and $W^s_{2,2}$. In our notation, it is equivalent to $\C_{1,1}=\C_{1,-1}=0$ and $\C_{2,2}=\C_{2,-2}$.

One notes that by replacing $r$ with $\chi r$ we have $\C_{m,n} \to \chi^m \, \C_{m,n}$. In other words, the order of $\C_{m,n}$ is indicated by $m$ in polar coordinates. Referring to Eq.~\eqref{CmnConst}, we find that there are $m+1$ cumulants with order $m$ in polar coordinates. As a result, we are able to find a one-to-one relation between $\C_{m, n}$ and $\A_{k\ell}$ with same order by equating two sides of the equation
\begin{equation}
\begin{aligned}
\sum_{m,n}&\frac{\A_{mn} (ik)^{m+n}\cos^m\varphi_k \sin^n\varphi_k }{m!n!}=\\
&\hspace*{1cm}\sum_{m=0}^{\infty}\sum_{n=-\infty}^{\infty} \frac{\C_{m,n}\; (ik)^m e^{in\varphi_k}}{m!}.
\end{aligned}
\end{equation}
In the left-hand side of the above equality, we re-write $\sum_{m,n}((ik_x)^m (ik_y)^n \A_{mn}/(m!n!))$ in polar coordinates. A few first  cumulants in polar coordinates can be written in terms of $\A_{mn}$ as follows,
\begin{subequations}\label{AA5}
	\begin{eqnarray}
	\C_{1,1}&=&\frac{1}{2}\left[\A_{10}-i\A_{01}\right], \\
	\C_{2,0}&=& \frac{1}{2}\left[\A_{20}+\A_{02}\right],\label{A5b}\\
	\C_{2, 2}&=&\frac{1}{4}\left[\A_{20}-\A_{02}-2i\A_{11}\right], \\
	\C_{3, 1}&=&\frac{3}{8}\left[\A_{30}+\A_{12}-i(\A_{03}+\A_{21})\right],\\
	\C_{3,3}&=&\frac{1}{8}\left[\A_{30}-3\A_{12}+i(\A_{03}-3\A_{21})\right].
	\end{eqnarray}
\end{subequations}
The cumulants $\C_{m,-n}$ can be obtained easily by using the reality condition. One can invert the above equations to find $\A_{mn}$ in terms of $\C_{m,n}$ too,
\begin{subequations}\label{CartToPolarCumul}
	\begin{eqnarray}
	\A_{10}&=&\C_{1,1}+\C_{1,-1},\\
	\A_{01}&=&i(\C_{1,1}-\C_{1,-1}),\\
	\A_{20}&=&\C_{2,0}+\C_{2,2}+\C_{2,-2},\\
	\A_{11}&=&i(\C_{2,2}-\C_{2,-2}),\\
	\A_{02}&=&\C_{2,0}-\C_{2,2}-\C_{2,-2},\\
	\A_{30}&=&\C_{3,3}+\C_{3,1}+\C_{3,-1}+\C_{3,-3},\\
	\A_{21}&=&\frac{i}{3}\left(3\C_{3,3}+\C_{3,1}-\C_{3,-1}-3\C_{3,-3}\right),\\
	\A_{12}&=&-\frac{1}{3}\left(3\C_{3,3}-\C_{3,1}-\C_{3,-1}+3\C_{3,-3}\right),\\
	\A_{03}&=&-i\left(\C_{3,3}-\C_{3,1}+\C_{3,-1}-\C_{3,-3}\right).
	\end{eqnarray}
\end{subequations}

For rotationally symmetric distributions, only the moments $\la r^{2m} \ra $ survive in the polar coordinates. In such a case, the only non-zero polar cumulants are $\C_{2k,0}$. For instance, all the cumulants in Eq.~\eqref{AA5} vanish except $\C_{2,0}$ where it is equal to $\la r^2 \ra /2$. A few first $\C_{2k,0}$ are given by
\begin{subequations}\label{PolarCumulSymm}
	\begin{eqnarray}
	\C_{2,0}&=&\frac{1}{2}\left[\la r^2 \ra\right], \\
	\C_{4,0}&=&\frac{3}{8}\left[\la r^4 \ra-2\la r^2 \ra^2\right], \\
	\C_{6,0}&=&\frac{5}{16}\left[\la r^6 \ra-9\la r^4 \ra\la r^2 \ra+12\la r^2 \ra^3\right], \\
	\C_{8,0}&=&\frac{35}{128}\left[\la r^8 \ra-16\la r^6 \ra\la r^2 \ra-18\la r^4 \ra^2\right.\nn\\
	&&\hspace*{1.5cm} \left.+144\la r^4 \ra\la r^2\ra^2-144\la r^2 \ra^4 \right]. 
	\end{eqnarray}
\end{subequations}
By replacing $r$  with $v_n$ in the Eq.~\eqref{PolarCumulSymm} and comparing it with Eq.~\eqref{1DnPartCumul}, we find that  $c_n\{2k\}\propto \C_{2k,0}$.

\section{Gram-Charlier A Series}\label{AppB}
\subsection{Gram-Charlier A Series: One Dimension}\label{AppBI}

In Sec.~\ref{IIIA}, we have introduced an iterative method to find the Gram-Charlier A series for a one-dimensional distribution. Also, we shortly explained the standard method of finding the series by using the orthogonal polynomials. Here, we use another standard method of finding the Gram-Charlier A series.

Let us consider the Characteristic Function of a given distribution $p(x)$ as $G(t)=\int dx\, e^{i t x} p(x) $. Then one can find the cumulants of the distribution by using Cumulative Function,
\bea\label{characCumul}
\log G(t)=\sum_{n=1}\kappa_n (i t)^n/n!.
\eea
Now assume that $p(x)$ can be approximated by a known distribution, namely a Gaussian distribution, 
$$\mathcal{N}(x)=\frac{1}{\sqrt{2\pi}\sigma}e^{-\frac{(x-\mu'_1)^2}{2\sigma^2}}.$$ 
By setting the mean value $\mu'=0$ and the width $\sigma=1$  for simplicity, the generating function of a Gaussian distribution can be simply find as $G_{\N}(t)=e^{-t^2/2}$. Referring to Eq.~\eqref{characCumul}, we have $G(t)=e^{\sum_{n=1}\kappa_n (i t)^n/n!}$. Additionally, we consider that the mean value and the variance of the distribution $p(x)$ are $\kappa_1=\mu'=0$ and $\kappa_2=\sigma^2=1$.  It means that we are able to write $G(t)$ for a generic distribution $p(x)$ as follows, 
\begin{equation}\label{1Dcharac}
\begin{aligned}
G(t)&=e^{\sum_{n=3}\kappa_n (i t)^n/n!}e^{-t^2/2}\\
&= e^{\sum_{n=3}\kappa_n (i t)^n/n!}G_{\N}(t).
\end{aligned}
\end{equation}
Now, we are able to find $p(x)$ by using inverse Fourier transformation. By considering the relation $\int dt (i t)^n e^{-i t x}$ $\times e^{-t^2/2}=(-d/dx)^n \int dt e^{-i t x}e^{-t^2/2}$, we have
\begin{equation}
\begin{aligned}
p(x) &=e^{\sum_{n=3}\kappa_n (-d/dx)^n/n!} \left[\int dt e^{-i t x}e^{-t^2/2}\right]/2\pi \\
&=e^{\sum_{n=3}\kappa_n (-d/dx)^n/n!}\left[e^{-x^2/2}/\sqrt{2\pi}\right].
\end{aligned}
\end{equation}
By scaling $x\to x/\sigma$ and shifting $x\to x-\mu'$, we find\footnote{Note that after scaling $x\to x/\sigma$, we have $(d/dx)\to \sigma (d/dx)$. Additionally, $\kappa_r$ scales to $\kappa_r/\sigma^r$. These two scalings cancel each other and we find Eq.~\eqref{B3}. }
\bea\label{B3}
p(x)=e^{\sum_{n=3}\kappa_n (-d/dx)^n/n!}\N(x).
\eea
Recall that the probabilistic Hermite polynomial is defined as $He_r(x)=e^{x^2/2}(-d/dx)^r e^{-x^2/2}$. As a result, one can simply find 
\bea\label{hermitEigen}
\left(-\frac{d}{dx}\right)^n \N(x)=\frac{1}{\sigma^n} He_n((x-\mu')/\sigma) \N(x).
\eea

It is worth noting that the Eq.~\eqref{B3} is exact. We are able to approximate this exact relation by expanding it in terms of number of derivatives. The result of such an expansion is called Gram-Charlier A series,
\begin{equation}
\begin{aligned}
p(x)=\frac{1}{\sqrt{2\pi}\sigma} &e^{-\frac{(x-\kappa_1)^2}{2\sigma^2}}\sum_{n=1} \frac{h_n}{n!} He_n((x-\kappa_1)/\sigma),\nn
\end{aligned}
\end{equation}
where $h_n$ has presented in Eq.~\eqref{A5} ($h_1=1,h_2=0$).

\subsection{Gram-Charlier A Series: Two Dimensions}\label{AppBII}

Consider the following two dimensional Gaussian distribution,
\bea\label{2DGaussian}
\N(\br)=\frac{1}{2\pi \sigma_x \sigma_y} e^{-\frac{(x-\mu'_x)^2}{2\sigma_x^2}-\frac{(y-\mu'_y)^2}{2\sigma_y^2}},
\eea
which is more general than Eq.~\eqref{2DGauss}.  Referring to Eq.~\eqref{generatingFu}, we can find the characteristic function of  Eq.~\eqref{2DGaussian} as follows
\bea
G_{\N}(\bk)=e^{-\frac{1}{2}(k_x \sigma_x+k_y \sigma_y)}.
\eea
Now, assume that the mean values $\la x \ra$ and $\la y \ra$ ($\la \cdots \ra =\int dx dy \cdots p(\br)$) are exactly equal to $\mu'_x$ and $\mu'_y$, respectively, and alternatively set $\mu'_x=\mu'_y=0$  by a shift. Then, the characteristic function of $p(\br)$ can be written as follows
\begin{equation}\label{B7}
\begin{aligned}
G(\bk)&=\exp{\sum_{m,n}\frac{(i\,k_x)^m (i\,k_y)^n \A_{mn}}{m! n!}}\\
&=\left[\exp{\sum_{m,n}\frac{(i\,k_x)^m (i\,k_y)^n \A'_{mn}}{m! n!}}\right] G_{\N}(\bk),
\end{aligned}
\end{equation}
where in the last line $\A'_{mn}$ are exactly as $\A_{mn}$ except 
\bea\label{Aprim}
\A'_{20}=\A_{20}-\sigma_x,\qquad \A'_{20}=\A_{02}-\sigma_y.
\eea
Note that we could choose $\A_{20}=\sigma_x$ and $\A_{02}=\sigma_y$ similar to what we did in one dimension. But here we do not use this convention for the future purpose.

We can find $p(\br)$ by an inverse Fourier transformation,
\bea\label{B9}
p(\br)=\left[\exp{\sum_{m,n}\frac{(-1)^{m+n} \A'_{mn}}{m! n!}}\left(\frac{d^{m+n}}{dx^m dy^n}\right)\right] \N(\br),\qquad 
\eea
where for finding the above we replaced $(i k_x)^m(i k_x)^n$ with $(-1)^{m+n}(d^{m+n}/dx^m dy^n)$ in Eq.~\eqref{B7}.

The Gram-Charlier A series can be found by expanding the exponential in Eq.~\eqref{B9} in terms of number of derivatives. By recovering  $\mu'_x$ and $\mu'_y$ by a reverse shift in $\br$,
we find 
\bea\label{2DGramm}
\hspace*{-0.6cm}p(\br)=\N(\br)\sum_{\substack{m,n=0}}\frac{h_{mn}}{m!n!} He_m(\frac{x-\mu_x}{\sigma_x})He_n(\frac{y-\mu_y}{\sigma_y}).\;\quad 
\eea
In the above, we have used a trivial extension of Eq.~\eqref{hermitEigen} to two dimensions. The coefficients $h_{mn}$ for $m+n\leq 2$ in the Eq.~\eqref{2DGramm} are given as follows (we set $\A_{11}=0$ for simplicity):
\begin{equation}\label{hmnTwo}
\begin{aligned}
h_{00}&= 1,\\
h_{10}&=h_{01}=0,\\
h_{20}&=\frac{\A_{20}}{\sigma_x^2}-1,\;  h_{11}=0,\; h_{02}=\frac{\A_{02}}{\sigma_y^2}-1.
\end{aligned}
\end{equation} 
Additionally, we have
\bea\label{hmnTwoII}
h_{mn}=\frac{\A_{mn}}{\sigma_x^m\sigma_y^n},\qquad 3\leq m+n\leq 5.
\eea
For $m+n=6$ we have,
\begin{equation}\label{hmnSix}
\begin{aligned}
h_{60}&=\frac{1}{\sigma_x^6}\left[\A_{60}+10\A_{30}^2\right],\\
h_{51}&=\frac{1}{\sigma_x^5\sigma_y}\left[\A_{51}+10\A_{30}\A_{21}\right],\\
h_{42}&=\frac{1}{\sigma_x^4\sigma_y^2}\left[\A_{42}+4\A_{30}\A_{12}+6\A_{21}^2\right],\\
h_{33}&=\frac{1}{\sigma_x^3\sigma_y^3}\left[\A_{33}+\A_{30}\A_{03}+9\A_{12}\A_{21}\right],\\
h_{24}&=\frac{1}{\sigma_x^2\sigma_y^4}\left[\A_{24}+4\A_{03}\A_{21}+6\A_{12}^2\right],\\
h_{15}&=\frac{1}{\sigma_x\sigma_y^5}\left[\A_{15}+10\A_{03}\A_{12}\right],\\
h_{06}&=\frac{1}{\sigma_y^6}\left[\A_{06}+10\A_{03}^2\right].\\
\end{aligned}
\end{equation} 
The other $h_{mn}$ can be found accordingly.

\subsection{Gram-Charlier A Series and Energy Density Expansion}\label{AppBIII}

We should say that the Gram-Charlier A series (Eq.~\eqref{2DGramm}) is exactly what has been used in Ref.~\cite{Teaney:2010vd} to quantify the initial energy density deviation from a rotationally symmetric Gaussian distribution\footnote{If we had considered $\A_{20}=\sigma_x$ and $\A_{02}=\sigma_y$ in Eq.~\eqref{Aprim}, we would have found the energy density distribution around a Gaussian distribution which is not rotationally symmetric.}. 

In Ref.~\cite{Teaney:2010vd}, the cumulants of the energy density $\rho(\mathbf{r})$ have been expanded in terms of moments
\bea\label{polarMoments}
\hat{\ve}_{m,n}\equiv \ve_{m,n}e^{i n\Phi_{m,n}}=-\frac{\{r^m e^{in\varphi}\}}{\{r^m\}}, 
\eea
where $m,n=0,1,\ldots\;$ and 
\[
\{\cdots\}=\frac{\int \cdots \rho(\mathbf{r})dxdy}{\int \rho(\mathbf{r}) dxdy}.
\]

Due to the translational invariance, we can freely choose the origin of the coordinates such that $\{r e^{\pm i \varphi}\}=0$. Using this and referring to  Eq.~\eqref{polarCumulantExampl} and Eq.~\eqref{polarMoments},  one finds 
\begin{equation}\label{polarCumulantExamp2}
\begin{aligned}
&\C_{2,0}=\frac{1}{2}\{ r^2\}, \\
&\C_{2, 2}=-\frac{1}{4}\{ r^2\}\ve_{2,2}e^{-2i\Phi_{2,2}}, \\
&\C_{3, 1}=-\frac{3}{8}\{ r^3\}\ve_{3,1}e^{-3i\Phi_{3,1}},\\
&\C_{3,3}=-\frac{1}{8}\{ r^3\}\ve_{3,3}e^{-3i\Phi_{3,3}}.\\
\end{aligned}
\end{equation}
It is common in the literature to define initial anisotropies as $\ve_1\equiv \ve_{3,1},\; \ve_2\equiv \ve_{2,2}$ and $\ve_3\equiv \ve_{3,3}$. According to Eq.~\eqref{polarCumulantExamp2}, initial anisotropies are itself cumulants. In fact, the parameter $\ve_n$ quantifies the global features of the initial energy density. For instance, $\ve_1$, $\ve_2$ and $\ve_3$ quantify dipole asymmetry, ellipticity and triangularity of the distribution respectively. The symmetry planes $\Phi_1\equiv \Phi_{3,1},\; \Phi_2$ $\equiv \Phi_{2,2}$ and $\Phi_3\equiv \Phi_{3,3}$ quantify the orientation of the dipole asymmetry, ellipticity and triangularity with respect to a reference direction. 

Furthermore, one can relate two dimensional cumulants in Cartesian coordinates, $\A_{mn}$, to those in polar coordinates, $\C_{m,n}$ by using Eq.~\eqref{CartToPolarCumul}. Using them, we can find
\begin{equation}\label{CartesCumulantExamp}
\begin{aligned}
&\A_{20}=\frac{1}{2}\{r^2\}(1-\ve_2\cos 2\Phi_2 ),\\ 
&\A_{11}=-\frac{1}{2}\{r^2\}\ve_2\sin 2\Phi_2 , \\
&\A_{02}=\frac{1}{2}\{r^2\}(1+\ve_2\cos 2\Phi_2 ), \\
&\A_{30}=-\frac{1}{4}\{ r^3\}(3\ve_1\cos \Phi_1+\ve_3\cos 3\Phi_3), \\
&\A_{21}=-\frac{1}{4}\{ r^3\}(\ve_1\sin \Phi_1+\ve_3\sin 3\Phi_3),\\
&\A_{12}=-\frac{1}{4}\{ r^3\}(\ve_1\cos \Phi_1-\ve_3\cos 3\Phi_3),\\
&\A_{03}=-\frac{1}{4}\{ r^3\}(3\ve_1\sin \Phi_1-\ve_3\sin 3\Phi_3).
\end{aligned}
\end{equation}

Now, we use Eq.~\eqref{2DGramm} to find a Gram-Charlier A series for the energy density. The result is the following,
\begin{equation}\label{teanyGram}
\begin{aligned}
\rho(r,\varphi) \propto e^{-\frac{r^2}{\{r^2\}}}\left[1+\sum_{n=1}^{3} a_n(r)\ve_n\cos n(\varphi-\Phi_n)\right],
\end{aligned}
\end{equation}
where
\begin{equation}\label{aOfr}
\begin{aligned}
a_1(r)&=-\frac{r\{r^3\}(r^2-2\{r^2\})}{\{r^2\}^3},\\
a_2(r)&=-\frac{r^2}{\{r^2\}},\\
a_3(r)&=-\frac{r^3\{r^3\}}{3\{r^2\}^3}.
\end{aligned}
\end{equation}
In the above, we assumed that $\sigma_x^2=\sigma_y^2=\{r^2\}/2$. This result coincides with that has been presented in Ref.~\cite{Teaney:2010vd}. Cumulants $\C_{m,n}$, $m>3$ cannot be written simply in terms of one moment $\ve_{m,n}$, however, it is always possible to find Gram-Charlier A series Eq.~\eqref{teanyGram} to any order beyond triangularity.

\section{A Radial Distribution From 2D Gram-Charlier A Series}\label{AppC}

The Bessel-Gaussian distribution \eqref{BG} was obtained by averaging out the azimuthal direction of a two dimensional Gaussian distribution \eqref{2DGauss}. Alternatively, one can do the same calculation and find a one dimensional series by averaging out the azimuthal direction of Eq.~\eqref{2DGramm}. We should emphasize that this calculation has already been done in Ref.~\cite{Abbasi:2017ajp} for the case that $p(\br)$ is rotationally symmetric around the origin. However, we will try to find such a radial series for a more general case.

For simplicity, we assume that $\mu'_y\equiv\A_{01}=0$ ($\mu'_x\equiv\A_{10}$) together with $\sigma\equiv\sigma_x=\sigma_y=\A_{20}=\A_{02}$. Also, we consider that $\A_{11}=0$. These considerations are compatible with $p(v_{n,x},v_{n,y})$. For such case, $h_{20}=h_{11}=h_{02}=0$.

By averaging out the azimuthal direction of Eq.~\eqref{2DGramm}, we will find a distribution which has the following form,
\begin{equation}\label{2DGramtoOne}
\begin{aligned}
&p(r)=\left(\frac{r}{\sigma^2}\right)e^{-\frac{r^2+\mu'^2_x}{2\sigma^2}} \left[I_0\left(\frac{r \mu'_x}{\sigma^2}\right)\sum_{i=1}P_{1,i}(r)\right.\\ &\hspace*{2.3cm}\left.+\left(\frac{r\mu'_x}{2\sigma^2}\right) I_1\left(\frac{r \mu'_x}{\sigma^2}\right)\sum_{i=1} P_{2,i}(r)\right].
\end{aligned}
\end{equation}
We can find the polynomials $P_{1,i}(r)$ and $P_{2,i}(r)$ by direct calculations. For $i=1,2$, one simply finds that $P_{1,1}(r)=1$ and $P_{2,1}(r)=P_{1,2}(r)=P_{2,2}(r)=0$. The polynomials for $i=3$ are given by,
\begin{equation}\label{P13}
\begin{aligned}
P_{1,3}(r)&=\frac{\mu'_x}{2\sigma}h_{12}+\frac{\mu'_x}{2\sigma}\left(1-\frac{\mu'^2_x}{3\sigma^2}\right)h_{30}\\
&+\frac{1}{6}\left[3h_{12}-\left(\frac{3\mu'^2_x}{\sigma^2}+1\right)h_{30}\right]\left(\frac{r^2}{\mu'_x\sigma}\right),\\
P_{2,3}(r)&=-\frac{2\sigma}{\mu'_x}\left(1+\frac{\sigma^2}{\mu'^2_x}\right)h_{12}\\
&+\frac{\mu'_x}{\sigma}\left(1+\frac{2\sigma^4}{3\mu'^4_x}\right)h_{30}+\frac{h_{30}}{3}\left(\frac{r^2}{\mu'_x\sigma}\right).
\end{aligned}
\end{equation} 
Finding $P_{1,4}(r)$ and $P_{2,4}(r)$ are straightforward, but they are more complicated. For that reason, we decided not to present them considering the limited space. Note that at each $P_{a,i}(r)$ only cumulants $\A_{mn}$ (if there is any) with order $m+n=i$ are present\footnote{Note that $\A_{mn}=\sigma^{m+n}h_{mn}$ for $m+n=3,4$ based on our assumptions in this section.}.

In Sec.~\ref{IIA}, we have argued that the one-dimensional distributions like $p(v_n)$ can be characterized by $c_n\{2k\}$. As a result, we expect to relate $\A_{mn}$ to $c_n\{2k\}$ by considering Eq.~\eqref{2DGramtoOne} (replace $r$ with $v_n$ and $\mu'_x$ with $\vb_n$). Considering Eq.~\eqref{1DnPartCumul} and using Eq.~\eqref{2DGramtoOne} up to $i=4$, we can find $c_n\{2k\}$ for $k=1,2,3$ as follows,
\begin{subequations}\label{2dCumulcheck}
	\begin{eqnarray}
	c_n\{2\}&=&\vb_n^2+2\sigma^2,\label{2dCumulcheckA}\\
	c_n\{4\}&=&-\vb_n^4+4\vb_n\left(\A_{12}+\A_{30}\right)\nn\label{2dCumulcheckB}\\
	&&\hspace*{1.5cm}+\A_{40}+2\A_{22}+\A_{04},\\
	c_n\{6\}&=&4\vb_n^6-8\vb_n^3(3\A_{12}+2\A_{30})\nn\\
	&&\hspace*{2cm}+6\vb_n^2(\A_{40}-\A_{04}).
	\end{eqnarray}
\end{subequations}

The above result is not surprising. In fact, it is a general relation between $\A_{mn}$ of the distribution $p(v_{n,x},v_{n,y})$ and $c_n\{2k\}$ of $p(v_n)$. Recall the equality $\la  e^{i v_n k\cos(\varphi-\varphi_k)} \ra_{\text{2D}}$ $=\la J_0(k v_n) \ra_{\text{1D}}$ in Eq.~\eqref{2DCharto1DChar} where the average $\la \cdots \ra_{\text{2D}}$ is performed with respect to the distribution $\tilde{p}(v_{n,x},v_{n,y})$. For the case that the average is performed with respect to $p(v_{n,x},v_{n,y})$, this relation should be replaced by
\bea
\frac{1}{2\pi}\int_{0}^{2\pi}d\varphi_k\,\la  e^{i v_n k\cos(\varphi-\varphi_k)} \ra_{\text{2D}}=\la J_0(k v_n) \ra_{\text{1D}}.
\eea
Using above, Eq.~\eqref{2DCumuls} and Eq.~\eqref{cnCumul}, we find
\bea
\log \frac{1}{2\pi}\int_{0}^{2\pi}d\varphi_k\,&&\exp \sum_{m,n}\frac{\A_{mn} (ik)^{m+n} }{m!n!}\left(\cos^m\varphi_k \sin^n\varphi_k\right)\nn\\
&&=\sum_{m=1}\frac{(ik)^{2m} c_n\{2m\}}{4^m(m!)^2}.\nn
\eea
By expanding both sides in terms of $k$ and comparing the coefficients, one finds the relation between $c_n\{2k\}$ and $\A_{mn}$. The result for $c_n\{2\}$ and $c_n\{4\}$ are as follows
\begin{subequations}\label{cn2ToAmn}
	\begin{eqnarray}
	c_n\{2\}&=&\A_{01}^2+\A_{10}^2+\A_{02}+\A_{20},\label{cn2ToAmnA}\\
	c_n\{4\}&=&-\A_{01}^4-2 \A_{10}^2 \A_{01}^2+2 \A_{02} \A_{01}^2-2 \A_{20} \A_{01}^2\label{cn2ToAmnB}\\
	&+&4 \A_{03} \A_{01}
	+8 \A_{10} \A_{11} \A_{01}+4 \A_{21} \A_{01}-\A_{10}^4+\A_{02}^2\nn\\
	&-&2 \A_{02} \A_{10}^2+4 \A_{11}^2  
	+\A_{20}^2+\A_{04}+4 \A_{10} \A_{12}\nn\\
	&+&2 \A_{10}^2 \A_{20}-2 \A_{02} \A_{20}+2 \A_{22}
	+4 \A_{10} \A_{30}+\A_{40}.\nn
	\end{eqnarray}
\end{subequations}
The relation for $c_n\{6\}$ is more complicated. Note that by setting the simplifications we used in finding  Eq.~\eqref{2DGramtoOne}, the above relations reduce to Eq.~\eqref{2dCumulcheckA} and Eq.~\eqref{2dCumulcheckB}.

In general, $c_n\{2k\}$ is written in terms of a large number of $\A_{pq}$, therefore, we are not able to write $\A_{pq}$ in terms of cumulants $c_n\{2k\}$.
However, it is possible to ignore the cumulants $\A_{pq}$ with $p+q\geq 4$ at some cases. For $n=2$, this truncation leads to a reasonable approximation for $c_2\{4\}$ and $c_2\{6\}$, and we are able to find $\A_{30}$ in terms of $c_2\{2\}$, $c_2\{4\}$ and $c_2\{6\}$ which has been done in Ref.~\cite{Giacalone:2016eyu}. In Ref.~\cite{Abbasi:2017ajp}, however, it has been shown that keeping 2D Cartesian cumulants up to forth order does not lead to a reasonable approximation for $c_2\{8\}$. Therefore, we are not able to find forth order $\A_{mn}$ in terms of $c_2\{2k\}$ simply.

Referring to Eq.~\eqref{GGCD2}, we see that the coefficients $\ell_{2i}$ (as a function of $c_n\{2k\}$ and $\vb_n$) appeared as an overall factors at each order of approximation. In contrast, the coefficients $h_{mn}$ in Eq.~\eqref{2DGramtoOne} do not appear in the polynomial $P_{a,i}(r)$ as an overall factor. Now, let us compute the limit $\mu'_x\to 0$. 
A straightforward calculation shows that
\bea
I_0\left(\frac{r \mu'_x}{\sigma^2}\right)P_{1,3}(r)+\left(\frac{r\mu'_x}{2\sigma^2}\right) I_1\left(\frac{r \mu'_x}{\sigma^2}\right) P_{2,3}(r)\to 0.\nn
\eea
As a result, the bracket in Eq.~\eqref{2DGramtoOne} is equal to one up to $i=3$. 

However, the bracket for $i=4$ is non-trivial in the limit $\mu'_x\to 0$,
\begin{equation}\label{i4Brack}\nn
\begin{aligned}
&I_0\left(\frac{r \mu'_x}{\sigma^2}\right)P_{1,4}(r)+\left(\frac{r\mu'_x}{2\sigma^2}\right) I_1\left(\frac{r \mu'_x}{\sigma^2}\right) P_{2,4}(r)\\
&\to\frac{1}{8} (h_{40}+2h_{22}+h_{04})\left(1-\frac{r^2}{\sigma^2}+\frac{r^4}{8\sigma^4}\right).
\end{aligned}
\end{equation} 
The last line in the above can be written as 
\bea
\frac{1}{8} (h_{40}+2h_{22}+h_{04}) L_2\left(\frac{r^2}{2\sigma^2}\right).
\eea
This result is interesting because it shows that the distribution \eqref{2DGramtoOne} has simpler form only if we assume $\mu'_x\equiv \A_{10}\to 0$. Let us again consider the flow harmonic distribution ($r \to v_n $ and $\mu'_x \to \vb_n$). By setting $\A_{20}=\A_{02}=\sigma^2$ and $\A_{10}=\A_{01}=\A_{11}=0$, we find from Eq.~\eqref{cn2ToAmnB} that 
\bea
c_n\{2\}&=& 2\sigma^2,\nn\\
c_n\{4\}&=&\A_{40}+2\A_{22}+\A_{04}.\nn
\eea
Therefore, the explicit form of Eq.~\eqref{2DGramtoOne} up to $i=4$ is given by
\begin{equation}\label{2DGramtoOneII}
\begin{aligned}
\hspace*{-0.2cm}p(v_n)=\left(\frac{v_n}{\sigma^2}\right)e^{-\frac{v_n^2}{2\sigma^2}} \left[1+\left(\frac{1}{2}\frac{c_n\{4\}}{c_n^2\{2\}}\right) L_2\left(\frac{v_n^2}{2\sigma^2}\right)\right],
\end{aligned}
\end{equation}
which is compatible with Eq.~\eqref{RGCODD} and what has been found in Ref.~\cite{Abbasi:2017ajp}. We can check that for $\A_{11}\neq 0$ the above result is true too\footnote{In this case, we have $c_n\{4\}=4\A_{11}^2+\A_{40}+2\A_{22}+\A_{04}$.}. 

It is worth noting that we have not assumed the original distribution to be rotationally symmetric. In fact,  the original distribution can have non-zero $\A_{30}$ or $\A_{12}$ too, but they are not appeared in $c_n\{2k\}$ or the distribution \eqref{2DGramtoOneII}. It is due to the fact that their information is lost during the averaging.

One could ask that while $h_{12}$ and $h_{30}$ are present in the polynomials $P_{1,3}(v_n)$ and $P_{2,3}(v_n)$ in Eq.~\eqref{P13}, it maybe possible to find all $h_{mn}$ from Eq.~\eqref{2DGramtoOne} by fitting it to an experimental $p(v_n)$. We should say that always a combination of $h_{mn}$ appears in the polynomials $P_{a,i}(v_n)$. As a result, if we assume that the series is convergent and also we have access to an accurate experimental distribution, we can only find a combination of $h_{mn}$ due to the information loss during the averaging. Specifically, in the simple Eq.~\eqref{2DGramtoOneII} case, only $\A_{40}+2\A_{22}+\A_{04}$ can be found by fitting.

In any case, the distribution \eqref{2DGramtoOne} is different from that mentioned in Eq.~\eqref{GGCD2}. These two different series refer to two different truncations. Note that $q_{n}\{2k\}$ can be written in terms of $\A_{pq}$ with $p+q\leq 2k$. For instance, if we keep $\A_{mn}$ up to forth order, all the cumulants $q_n\{2k\}$ are non-zero. Therefore, in principal, the summation \eqref{GGCD2} goes to infinity while the summation \eqref{2DGramtoOne} is truncated up to $i=4$.

\section{RGC Distribution From Multiple Orthogonal Polynomials}\label{appD}

In Sec.~\ref{seIII}, we have discussed that how the orthogonality conditions of $He_n(x)$ and $L_n(x)$ can be used to find Eq.~\eqref{GCA} and Eq.~\eqref{RGCODD}, respectively. However, ordinary orthogonal polynomials cannot be employed for the more general distribution series \eqref{GGCD2}. Recall that in one- dimension the polynomials  $\PP_n(x)$ and  $\PP_m(x)$ are orthogonal with respect to the measure $w(x)dx$ if $\int \PP_n(x)\PP_m(x) w(x)$ $dx=\zeta_n\delta_{nm}$. This relation can be equivalently written as $\int x^m$ $\PP_n(x) w(x) dx=0$ for $0\leq m<n$ \cite{ismail}.

We will show we can obtain Eq.~\eqref{GGCD2} by employing a generalization of the orthogonal polynomials. These generalized orthogonal polynomials are called Multiple Orthogonal Polynomials (MOPs) \cite{ismail}. The MOPs are orthogonal with respect to more than one weight. Let us define $r$ different weights $\vec{w}(x)=(w_1(x),\ldots,w_r(x))$ and a multi-index $\vec{n}=(n_1,\ldots,n_r)$ ($n_i\in \mathbb{N}_0$). Then the \textit{type I multiple orthogonal polynomial} is given as $\vec{\PP}_{\vec{n}}=(\PP_{1,\vec{n}},\ldots,\PP_{r,\vec{n}})$ where $\PP_{j,\vec{n}}$ is a polynomial with degree at most $n_j$ \footnote{Another class of  MOPs are called \textit{type II multiple orthogonal polynomial}. In this class, a multi-indexed monic polynomial $\PP_{\vec{n}}$ satisfies the following $r$ orthogonality conditions,
	\[
	\int x^m \PP_{\vec{n}} w_i(x) dx,\qquad 0\leq m < n_i, \nn
	\] 
	where $0\leq i\leq r$ \cite{ismail}. Type II MOPs are not used here.}. 
These polynomials obey the following orthogonality condition,
\bea\label{MOP}
\int  x^m \vec{\PP}_{\vec{n}}\bigcdot \vec{w}(x)\,dx=0,\qquad 0\leq m < N+r-1, \quad
\eea
where $N=\sum_{i=1}^{r} n_i$, and dot indicates the inner product. Note that for $r=1$, the above relation returns to the ordinary orthogonal polynomial.

Here, we restrict ourselves to the \textit{weakly complete systems}. In this case, the multi-index $\vec{n}$ always has the following form \cite{Kalyagin,VanAssche},
\bea\label{nIndex}
(n_1,\ldots,n_r)=(\underbrace{k+1,\ldots k+1}_{j \text{ times}},\underbrace{k,\ldots k}_{r-j \text{ times}}),\quad k\in\mathbb{N}_0.\qquad 
\eea
Here $0\leq j<r$. In this case, we can define a one-to-one map between a single index $i$ and $\vec{n}$ as $i(\vec{n})=r k+j+1$. Consequently, we are able to define 
\bea\label{Qdefine}
\Y_{i(\vec{n})}(x)=\vec{\PP}_{\vec{n}}\bigcdot \vec{w}(x).
\eea
As a result, Eq.~\eqref{MOP} can be written as
\bea\label{MOPmodi}
\int dx \, x^m \Y_{i}(x)=0,\qquad  0\leq m<i.
\eea
Note that the above relation is \textit{not} an ordinary orthogonality condition because $\Y_i$ is not a single finite degree polynomial times a weight. 

An example of the MOPs can be found in Ref.~\cite{CoussementVanAssche} where the polynomials are orthogonal with respect to a weight vector $(w_{\nu,c}(x),w_{\nu+1,c}(x))$. Here, 
\bea
w_{\nu,c}(x)=x^{\nu/2}I_{\nu}(2\sqrt{x})e^{-c x},
\eea
is defined for $x>0$, $\nu>-1$ and $c>0$. In this study, we are interested in these specific MOPs because they are related to $\tilde{\Q}_i(v_n;\vb_n)$ in Eq.~\eqref{tildeQ}.

For the present specific case, we have $r=2$. As a result, the index $\vec{n}$ has the form $(k,k)$ or $(k+1,k)$ referring to Eq.~\eqref{nIndex}. The explicit form of a few first polynomials $\vec{\PP}_{\vec{n}}$ for $\nu=0$ are given by,
\begin{equation}\label{MOPExamp1}
\begin{aligned}
\PP_{1,(0,0)}&=1,\\
\PP_{2,(0,0)}&=-c,\\
\PP_{1,(1,1)}&=1+3\left(1+\frac{2c}{3}\right)c^2 x,\\
\PP_{2,(1,1)}&=-3c\left(1+c+\frac{2c^2}{3}\right)-c^3 x.
\end{aligned}
\end{equation} 
and
\begin{equation}\label{MOPExamp2}
\begin{aligned}
\PP_{1,(1,0)}&=1+c^2 x,\\
\PP_{2,(1,0)}&=-2\left(1+\frac{c}{2}\right)c,\\
\PP_{1,(2,1)}&=1+6 c^2\left(1+\frac{4c}{3}+c^2\right)x+c^4 x^2,\\
\PP_{2,(2,1)}&=-4c\left(1+\frac{3c}{2}+2 c^2+\frac{3c^3}{2}\right)\\
&\hspace*{4cm}-4c^3(1+c)x.
\end{aligned}
\end{equation} 
We refer the interested reader to Ref.~\cite{CoussementVanAssche}  for the general form of $\vec{\PP}_{\vec{n}}$. In Ref.~\cite{CoussementVanAssche}, $\PP_{1,(n_1,n_2)}$ and $\PP_{2,(n_1,n_2)}$ have been shown by $A_{n_1,n_2}^{(\nu,c)}$ and $B_{n_1,n_2}^{(\nu,c)}$.

The function $\Y_i(x)$ in Eq.~\eqref{Qdefine} has the following simple form \cite{CoussementVanAssche},
\bea
\Y^{\nu,c}_{i}(x)=\sum_{k=0}^{i} \binom{i}{k}(-c)^k x^{(\nu+k)/2}I_{\nu+k}(2\sqrt{x})e^{-cx}.\qquad
\eea
Now, we can choose
\[\nu=0,\qquad c=\frac{2\sigma^2}{\vb_n^2},  \]
and change the coordinate 
$$x=\left(\frac{v_n\vb_n}{2\sigma^2}\right)^2.$$
Referring to Eq.~\eqref{tildeQ}, one finds
\bea
\Y_i^{0,\frac{2\sigma^2}{d\vb_n}}\left(\frac{v_n^2\vb_n^2}{4\sigma^4}\right)=\tilde{\Q}_i(v_n;\vb_n)e^{-(v_n^2/2\sigma^2)}.
\eea
Consequently, the orthogonality condition \eqref{MOPmodi} up to an irrelevant numerical factor can be written as
\begin{equation}\label{Qortho}
\begin{aligned}
\int_{0}^{\infty} dv_n\, v_n^{2m}\,\tilde{\Q}'_i(v_n;\vb_n)=0,\qquad 0\leq m<i,
\end{aligned}
\end{equation} 
where
\bea
\tilde{\Q}'_i(v_n;\vb_n)=\frac{v_n}{\sigma^2}  e^{-\frac{v_n^2+\vb_n^2}{2\sigma^2}} \tilde{\Q}_i(v_n;\vb_n).
\eea
As a result, we are able to write $p(v_n;\vb_n)$ as a series of  $\tilde{\Q}'_i(v_n;\vb_n)$,
\bea\label{GGCD2App}
p(v_n;\vb_n)=\sum_{i=0}^{\infty}\frac{(-1)^i\ell_{2i}}{i!} \tilde{\Q}'_i(v_n;\vb_n),\hspace*{0.6cm}
\eea
with unknown $\ell_{2i}$. Note that Eq.~\eqref{GGCD2App} is exactly the Eq.~\eqref{GGCD2}. 

Let us define the following quantity,
\bea\label{mpri}
m'_{ri}= \int d v_n\, v_n^{2r} \tilde{\Q}'_i(v_n;\vb_n).
\eea
Noting Eq.~\eqref{Qortho}, we find that $m'_{ri}=0$ for $0\leq r<i$. As a result, we are able to write the moments of $p(v_n;\vb_n)$ as follows,
\begin{equation}\label{recursI}
\begin{aligned}
\la v_n^{2r} \ra&=\int_{0}^{\infty}dv_n v_n^{2r} p(v_n;\vb_n)\\
&= \sum_{i=0}^{\infty}\frac{(-1)^i\ell_{2i}}{i!} \int_{0}^{\infty}dv_n\, v_n^{2r} \tilde{\Q}'_i(v_n;\vb_n)\\
&= \sum_{i=0}^{r}\frac{(-1)^i\ell_{2i}}{i!} m'_{ri}.
\end{aligned}
\end{equation} 
Using the above relation, we can find the following recurrence relation for $\ell_{2r}$,
\bea
\ell_{2r}=(-1)^{r}r!\left[\frac{\la v_n^{2r} \ra-\sum_{i=0}^{r-1}\frac{(-1)^i\ell_{2i}}{i!} m'_{ri}}{m'_{rr}}\right],
\eea
where $r\geq0$. 

Note that $m'_{ri}$ in Eq.~\eqref{mpri} is completely computable, for instance $m'_{0,0}=1$. Consequently, we have $\ell_0=1/m'_{0,0}$ for $r=0$ which leads to $\ell_0=1$. The next iteration is as follows
\bea
\ell_2&=&\frac{\ell_0 m'_{1,0}-\la v_n^2 \ra}{m'_{1,1}}\nn\\
&=& \frac{\la v_n^2 \ra-\vb_n^2-2\sigma^2}{2\sigma^2}.\nn
\eea
By considering the Eq.~\eqref{1DnPartCumulA} and Eq.~\eqref{cn2Approx}, we find $\ell_2=0$. Similarly, we have
\bea
\ell_4 &=&\frac{\la v_n^4 \ra-2 \la v_n^2 \ra^2-\vb_n^2}{(\la v_n^2 \ra -\vb_n^2)^2}\nn\\
&=& \frac{c_n\{4\}+\vb_n^2}{(c_n\{2\}-\vb_n^2)^2},
\eea
where in the last line we used Eq.~\eqref{1DnPartCumulA} and Eq.~\eqref{1DnPartCumulB}. One can continue this procedure to find all the values of $\ell_{2i}$ as mentioned in Eq.~\eqref{generL}.




\end{document}